\documentclass[12pt]{article}
\usepackage[a4paper,margin=1in]{geometry}
\usepackage{amsmath}
\usepackage{amssymb}
\usepackage{amsfonts}
\usepackage{array}
\usepackage{graphicx}
\usepackage{mathrsfs}
\usepackage{multirow}
\usepackage{siunitx}
\usepackage{hyperref}
\hypersetup{
	colorlinks=true,
	linkcolor=blue,
	filecolor=blue,      
	citecolor=blue,
	urlcolor=blue
}
\usepackage[table]{xcolor}

\begin{document}
\title{Bacterial glass transition}

\author{
	Martin~Maliet$^{1,2}$,
        Nicolas~Fix-Boulier$^{1,2}$,
	Ludovic~Berthier$^{3}$,
	Maxime~Deforet$^{1,2\ast}$\and
	\small$^{1}$Sorbonne Université, CNRS, Laboratoire Jean Perrin, LJP, 75005, Paris.\and
        \small$^{2}$Sorbonne Université, CNRS, Inserm, Institut de Biologie Paris-Seine, IBPS, 75005 Paris.\and
	\small$^{3}$Gulliver, UMR CNRS 7083, ESPCI Paris, PSL Research University, 75005 Paris.\and
	\small$^\ast$Corresponding author. Email: maxime.deforet@sorbonne-universite.fr\and
}

\maketitle
\begin{abstract} 
Bacterial assemblies exhibit rich collective behaviors that control their biological functions, making them a relevant object of study from an active matter physics perspective. Dense bacterial suspensions self-organize into distinct physical phases with intriguing dynamical properties. Here, we study dense two-dimensional films of swimming bacteria using advanced imaging techniques and machine learning. By varying density, we uncover a bacterial glass transition, a direct active matter analogue of equilibrium glass transitions in colloidal and molecular fluids. The transition is marked by a dramatic slowdown of dynamics with minimal structural change. Strong dynamic heterogeneity emerges in space and time, leading to an anomalous violation of the Stokes-Einstein relation and a growing dynamic correlation length, universally observed across five bacterial strains. Our results establish that bacterial colonies exhibit glassy dynamics, but their living, active nature gives them unique properties, paving the way for new research regarding how non-equilibrium physics impacts biology.
\end{abstract}

In infectious settings, bacteria often transition from motile planktonic cells to surface-attached communities known as biofilms, that are highly resilient to antibiotics and immune responses~\cite{flemming2016biofilms}. This process involves complex regulatory processes—including mechanosensing, signaling, and matrix production~\cite{webster2022power, wang2023accumulation, flemming2023biofilm}. Beyond these biochemical pathways, dense bacterial populations also behave as active physical systems: driven by cellular processes such as growth and motility, they exhibit collective dynamics that emerge from many-body interactions between the self-propelled cells. The formation and evolution of dense communities are therefore also shaped by physical mechanisms~\cite{be2019statistical, maier2021physical, wong2021roadmap, hallatschek2023proliferating}, among which crowding plays a central role by altering cell motility and organization~\cite{beroz2018verticalization, meacock2021bacteria}. At intermediate densities, hydrodynamic interactions among motile bacteria may give rise to active turbulence, producing chaotic flow patterns~\cite{dombrowski2004self, dunkel2013fluid, peng2021imaging, henshaw2023dynamic}. As density increases further, these assemblies undergo a fluid-to-solid transition~\cite{dell2018growing, sengupta2020microbial, copenhagen2021topological, zhang2021morphogenesis, dhar2022self}. Here, our goal is to fully elucidate the consequences of crowding and to provide an in-depth experimental study of the nature of the dynamic slow-down and dynamic arrest in bacterial colonies. 

Due to asynchronous division cycles, bacterial cells exhibit a broad size distribution, spanning from newly divided cells to those approaching division~\cite{taheri2015cell, deforet2015cell}. This inherent variability prevents crystallization and long-range order, causing dense bacterial colonies to form amorphous solid structures at high densities. We refer to this process as a {\it bacterial glass transition}, by analogy with equilibrium colloidal and molecular glass transitions~\cite{berthier2011theoretical}.

Active glass transitions have been observed in systems of self-driven particles, where dynamic arrest emerges in the absence of long-range order~\cite{janssen2019active, berthier2019glassy}. A central question is whether active glasses exhibit behaviors comparable to their equilibrium counterparts, but addressing it is experimentally challenging. Such studies require precise control over experimental parameters such as particle density and motility, along with high-quality, high-resolution datasets capable of capturing the hallmarks of glass transitions that require space-time resolution over large length scales and time scales. 

Experimental studies of glass transitions in synthetic active systems, such as phoretic colloids~\cite{klongvessa2019active, ghosh2024onset} and robotic cell mimics~\cite{arora2024shape}, have revealed some glassy behaviors but are often constrained by small system sizes and limited dynamical properties. In contrast, a wider range of biological systems—including cell tissues, ant aggregates, and cytoplasmic components (see \cite{sadhukhan2024perspective} for a review)—have been observed to exhibit some glassy characteristics, with bacterial systems among them~\cite{lama2024emergence, sugino2024non}. However, the complexity of biological systems makes experimental control challenging, and quantification remains difficult due to limitations in detecting and tracking individual components over multiple scales. 

Here, we successfully overcome these long-standing challenges. Exploiting the ability of bacteria to form large monolayers on agar gels, we collected an unprecedented dataset spanning multiple orders of magnitude in both space and time. Using advanced imaging techniques combined with machine learning, we tracked thousands of individual cells over a broad range of densities, fully capturing the gradual transition to an amorphous bacterial glass. This comprehensive dataset not only enables us to rigorously characterize the overall slowdown in cell dynamics but also reveals its emerging complex fluctuations in space and time, testing the universality of our results by studying five different bacterial strains. Our comprehensive work offers a unique opportunity to analyze glassy dynamics far from equilibrium in a well-characterized and controllable active biological system.

\section*{Crowded and disordered bacterial monolayers}

Monolayers naturally form at the edge of swarming colonies of \textit{Pseudomonas aeruginosa}, see Fig.~\ref{fig:Exp_design}(A), a motile rod-shaped bacterium that alternates forward and backward motion (\textit{run-reverse}). This movement is driven by its polar flagellum, that spins alternately in counterclockwise and clockwise directions. We developed an experimental setup to observe and analyze large monolayers within these colonies (see Methods for further details \cite{methods}). Unlike \textit{Bacillus subtilis}, a bacterium known for its pronounced flocking behavior~\cite{zhang2010collective, jeckel2019learning}, \textit{P.~aeruginosa} cells do not exhibit collective flocking, presumably due to the dispersion in their swimming speeds. While individual \textit{P.~aeruginosa} cells transiently align, activity and polydispersity prevent both long-range nematic and tetratic order~\cite{gonzalez2017clustering} (fig.~\ref{sifig:Orientation}). Due to the asynchronous division cycle, cell lengths are widely distributed (fig.~\ref{sifig:Length}). This is a key factor in the emergence of a two-dimensional amorphous phase at high surface fractions, $\phi \geq 0.65$. Using \textsc{Distnet2D}, a state-of-the-art deep learning-based segmentation and tracking tool that leverages temporal information~\cite{ollion2024distnet2d}, we reconstructed the shape and dynamic trajectories of all cells within a broad field of view, following thousands of cells simultaneously over long times (Figs.~\ref{fig:Exp_design}(B, C) and fig.~\ref{sifig:all_cells} for an illustration of the entire field of view). Having access to detailed positional and orientational data across a wide range of surface fractions for a large number of particles enables us to compile the robust dataset spanning multiple orders of magnitude in both space and time needed to study the transition to a bacterial glass phase.

\section*{Slowing down of orientational and translational dynamics}

\label{sec:Bulk_dynamics}

The bacterial monolayer did not exhibit strong structural change as the density increases (fig.~\ref{sifig:Spatial_structure}). In contrast, a clear dynamical slowdown was evident in the experimental movies (Movies~\ref{simov:WT_low}, \ref{simov:WT_mid} and \ref{simov:WT_high}). To quantify this, we computed time correlation functions characterizing orientational and translational motion of individual cells. For orientations, the correlation function $C_{\theta}(\tau)$ was calculated by following the evolution of the angle $\theta_i(t_0)$ between the body of cell $i$ at time $t_0$ and the horizontal axis:
\begin{equation}
\label{eq:Ctheta}
    C_{\theta}(\tau) = \langle \frac{1}{N} \sum_{i} \cos[\theta_i(t_0+\tau)-\theta_i(t_0) ] \rangle ,
\end{equation}
\noindent where brackets indicate an average over $t_0$. For translations, the persistence function $C_p(\tau)$ was calculated by discretizing space into a lattice of boxes of linear size equal to the average cell length, and tracking the fraction of cells remaining in their initial box after a delay $\tau$:
\begin{equation}
\label{eq:Cp}
    C_p(\tau) = \langle \frac{1}{N} \sum_i P_i(t_0, t_0 + \tau) \rangle  ,
\end{equation}
with $P_i(t_0, t_0+\tau)=1$ as long as cell $i$ remains in the box it occupies at $t_0$, and $P_i=0$ after that. Fig.~\ref{fig:Bulk_dynamics}(A, B) show both correlation functions for a range of surface fractions and demonstrate the gradual, but very pronounced, slowdown of dynamics as $\phi$ increases.

To quantify the observed slowdown, we extracted the respective relaxation times $\tau_\theta$ and $\tau_p$ from fitting the time decay of correlation functions to stretched exponential forms. The characteristic time for translation $\tau_p$ corresponds to the average time it takes bacterium to travel a distance equal to its body length, a metric analogous to the $\alpha$-relaxation time in glassy systems. Instead, $\tau_\theta$ quantifies the average time to reorient by an angle roughly equal to $\pi/3$. The surface fraction dependence of both correlation times is shown in Fig.~\ref{fig:Bulk_dynamics}(C). The weak evolution for $\phi = 0.45-0.6$ becomes much sharper when $\phi \geq 0.6$. As found for Brownian colloids~\cite{hunter2012physics}, this behaviour can be well described by an exponential divergence 
\begin{equation}   
\tau(\phi) = \tau_0 \exp \left( \frac{K \phi}{\phi_c - \phi} \right), 
\label{eq:VFT}
\end{equation}
mathematically analogous to the Vogel-Fulcher-Tamman law used for molecular fluids~\cite{berthier2011theoretical}. In Eq.~(\ref{eq:VFT}), $\tau_0$ describes the relaxation time in the dilute limit, $K$ plays the role of a glass fragility, and $\phi_c$ is the critical surface fraction where timescales diverge. We obtain very close values for both degrees of freedom, $\phi_{c,\theta} = 0.679 \pm 0.002 $ and $\phi_{c,p} = 0.678 \pm 0.002$. The raw data $\tau_p(\phi)$ and $\tau_{\theta}(\phi)$ for all strains are in fig.~\ref{sifig:tau_phi_all} and VFT fitting parameters in fig.~\ref{sifig:VFT_times}. The strong similarity between timescales is further demonstrated in the parametric plot in Fig.~\ref{fig:Bulk_dynamics}(D), where a relation $\tau_p \propto \tau_\theta$ is obeyed across all timescales for all tested bacterial strains (best fit is $\tau_p \sim \tau_\theta^{0.95}$). A final indicator of dynamic arrest directly focusing on real space motion is the mean-squared displacement (MSD)
\begin{equation}
    MSD(\phi,\tau) = \langle \frac{1}{N} \sum_i | \vec{r_i}(t_0+\tau) - \vec{r_i}(t_0) |^2 \rangle . 
\end{equation}
We followed its evolution with $\phi$ at fixed $\tau=0.375$~s: see 
fig.~\ref{sifig:MSD_phi_all} for the raw data and Fig.~\ref{fig:Bulk_dynamics}(E) for a compilation of fits to Eq.~(\ref{eq:VFT}). For \textit{P.~aeruginosa}, we get $\phi_{c,MSD} = 0.683 \pm 0.002$, again very close to $\phi_{c,p}$ and $\phi_{c,\theta}$. Within statistical errors, these data confirm a simultaneous global arrest of all degrees of freedom for all strains, in sharp contrast with the conclusions of Ref.~\cite{lama2024emergence}.

To understand the possible influence of single-cell motility on the glassy dynamics, we repeated our analysis for various mutants of \textit{P.~aeruginosa} (list in Table \ref{si_table:Strains}). While most of our results are for the wild type, which has a single polar flagellum enabling forward and backward swimming, we also examined mutants with a similar aspect ratio to the wild type (fig.~\ref{sifig:Length} shows a comparison of cell size distributions): \textit{cheR1} swims unidirectionally (Movie~\ref{simov:C1}), the multi-flagellated hyperswarmer mutant exhibits increased swimming speed, and the $\Delta$\textit{pilA} mutant lacks type IV pili -- appendages critical for cell-cell and cell-substrate interactions, as well as for cell twitching. For these mutants, glass transitions occurred at a comparable critical surface fraction, $\phi_{c,MSD} = 0.673-0.683$, see Fig.~\ref{fig:Bulk_dynamics}(F) and fig.~\ref{sifig:MSD_phi_all}. The more elongated hyperswarmer mutant, \textit{frik}, with an average aspect ratio 44\% greater than the wild type, showed a lower critical surface fraction ($\phi_{c,\theta} = 0.642$ $\pm$ 0.003, very close to $\phi_{c,p} = 0.641$ $\pm$ 0.002) and a higher fragility $K$ (fig.~\ref{sifig:VFT} and Movie~\ref{simov:Cl5}). Even for this more elongated cell, we do not observe the decoupling between orientational and translational glass transitions reported in some earlier studies of elongated colloids~\cite{zheng2011glass} and {\it Escherichia coli} bacteria~\cite{lama2024emergence}. Overall, these results show that the main features of the bacterial glass transition are not affected by details of single-cell motility, but its precise location is more sensitive to geometry than to motility. 

\section*{Emergence of dynamic heterogeneity}

\label{sec:Cell_heterogeneities}

\begin{figure}
\begin{center}
\includegraphics[width = \linewidth]{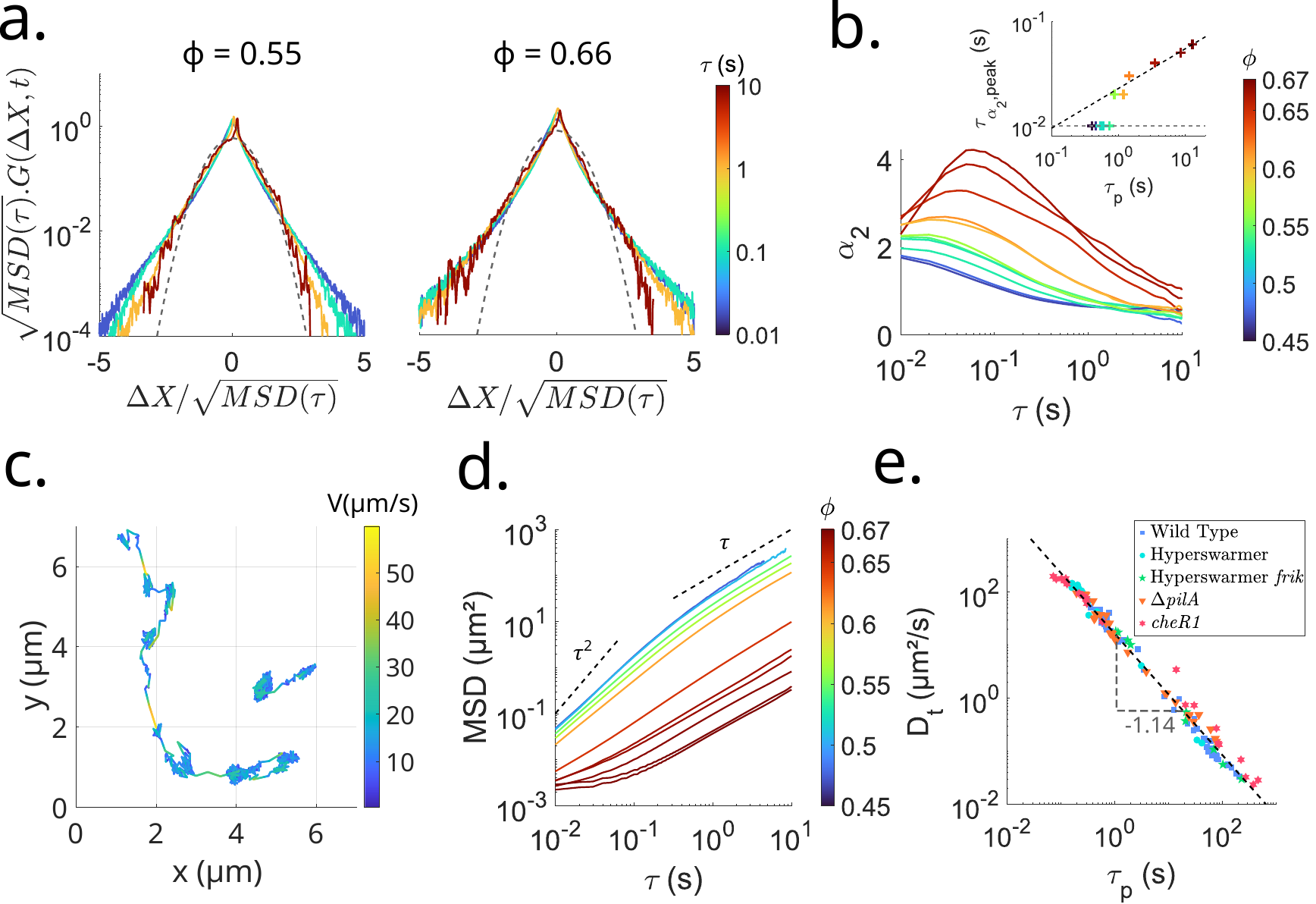}\\
\caption{\textbf{Emergent dynamic heterogeneity and anomalous decoupling.} 
(\textbf{A}) Van Hove distributions $P(\Delta X, \tau)$ at different times ($\tau=0.02$~s, 0.1~s, 1~s, 8~s) and two different surface fractions. Dashed line is the corresponding Gaussian distribution.  
(\textbf{B}) Time dependence of the non-Gaussian parameter, Eq.~(\ref{eq:NGP}), for a range of surface fractions. Inset: the time of the peak of $\alpha_2$ as a function of $\tau_p$. The gray horizontal dashed line is the lower limit due to image acquisition, the black dashed line is a power law of exponent 0.38, fitted on data points above the lower limit.  
(\textbf{C}) Two illustrative cell trajectories of the same duration (10~s) extracted in the same experiment at $\phi=0.661$, showing a fast moving cell performing several cage jumps, co-existing with a nearly arrested one. 
(\textbf{D}) Time dependence of the mean-squared displacements for a range of surface fractions. Dashed lines indicate ballistic ($\tau^2$) and diffusive ($\tau$) regimes. 
(\textbf{E}) Parametric evolution of the diffusion coefficient $D_t$ with the correlation time $\tau_p$ the wild-type strain and all mutants. The dashed line represents a fractional Stokes-Einstein decoupling with an anomalous exponent $\zeta \approx 1.14 > 1$.}
\label{fig:Cell_heterogeneities}
\end{center}
\end{figure}

Ensemble-averaged time correlation functions reveal a dramatic slowing down of the dynamics. For equilibrium fluids, this slowing down is accompanied by strong and specific fluctuations. These dynamic heterogeneities~\cite{berthier2011dynamical} physically imply the co-existence, at any moment, of fast and slow cell motion emerging from a broad distribution of dynamic behaviors. We analyzed the probability distribution of single bacteria displacements over different time delays, see Fig.~\ref{fig:Cell_heterogeneities}(A). This van Hove distribution is defined as $P(\Delta X,\tau) = \langle \frac{1}{N} \sum_i \delta [ \Delta X - \Delta X_i(\tau) ] \rangle$ with $\Delta X_i(\tau)$ the components of $\Delta \vec{r}_i(\tau)$ and $N$ the number of cells in the field of view. Distributions over a broad range of time delays and surface fractions reveal the existence of large-displacement tails extending much further than the corresponding Gaussian distribution. These near-exponential tails~\cite{chaudhuri2007universal} reveal the existence of a population of cells moving significantly faster than the average population, and, more broadly, of a displacement mechanism that differs qualitatively from Fickian diffusion. A Gaussian distribution is only slowly recovered at very large times, when the system eventually displays homogeneous diffusive motion. We quantify deviations from Gaussianity using the non-Gaussian parameter (NGP) $\alpha_2(\tau)$:
\begin{equation}
        \alpha_2(\tau) = \frac{1}{3} \frac{ \langle \Delta X(\tau)^4 \rangle}{ \langle \Delta X(\tau)^2 \rangle^2 } - 1,
\label{eq:NGP}
\end{equation}
which vanishes, by definition, when $P(\Delta X,\tau)$ is Gaussian. As shown in Fig.~\ref{fig:Cell_heterogeneities}(B), $\alpha_2(\tau)$ exhibits a growing maximum at an intermediate time increasing rapidly with $\phi$ (see inset). The growing maximum reveals an increasingly broad distribution of particle displacements as the dynamics slows down~\cite{kob1997dynamical}. In real space, the broad tails stem from a variety of individual trajectories, as illustrated in Fig.~\ref{fig:Cell_heterogeneities}(C). All cells are trapped (or, caged) over long periods of time, and undergo large jumps at widely distributed times. As a result, over a given observation time, some cells perform many jumps while others barely move. The emergence of transient caging is also revealed by the MSD shown in Fig.~\ref{fig:Cell_heterogeneities}(D). At low $\phi$, a ballistic to diffusive evolution typical of persistent random walks is observed. Instead, a pronounced sub-diffusive plateau regime appears at large density, directly reflecting caging, perhaps the most robust signature of glassy dynamics. 

\section*{Diffusion and anomalous Stokes-Einstein decoupling}

The long-time limit of the MSD defines the diffusion constant, $D_t = \lim_{t \to \infty} MSD / (4t)$ (see Methods for data analysis~\cite{methods}). In simple fluids, the Stokes-Einstein relation states that $D_t$ is inversely proportional to the viscosity. In supercooled liquids, the viscosity is proportional to the relaxation times (here, $\tau_p$ and $\tau_\theta$), and the Stokes-Einstein relation becomes $D_t \propto \tau_p^{-1}$. Near the glass transition of thermal fluids, the strong heterogeneity of particle displacements leads to violations of the Stokes-Einstein relation~\cite{ediger2000spatially}, or, more generally, to a decoupling between various transport properties, often taking the form of a fractional relation, $D_t \sim \tau_p^{-\zeta}$, with $0 < \zeta < 1$ an empirical exponent. Our results for dense bacteria are in  Fig.~\ref{fig:Cell_heterogeneities}(E), showing a parametric plot of $D_t$ against $\tau_p$ for the five strains studied. Remarkably, a strong decoupling is observed for all systems, that can be described by a unique exponent, $\zeta \approx 1.14$, independently of mutations and body geometry. In equilibrium systems, decoupling is explained by the fact that $D_t$ is dominated by the motion of fast particles while $\tau_p$ is controlled by the slow ones, thus leading to $\zeta < 1$~\cite{ediger2000spatially,jung2004excitation}. Despite the presence of heterogeneity in our systems, we find instead a stronger variation of $D_t$ leading to $\zeta > 1$. To our knowledge, no such value was observed in equilibrium systems. We hypothesize that this anomalous decoupling is specific to active glassy systems, where persistent self-propulsion may significantly affect the self-diffusion process. Further research, for instance using simulations of highly persistent particles~\cite{keta2022disordered}, should elucidate this remarkable finding. 

\section*{Growth of a dynamic correlation length}

The emergence of broad distributions of particle displacements provides no information about the coherence and spatial organization of cell motion. Intuitively, crowding implies that a cell cannot move significantly if its neighbors do not simultaneously rearrange, like a person in a packed subway who can only move if others make space. Spatial correlations of the dynamics are a hallmark of glassy dynamics~\cite{berthier2011dynamical}, and were also reported in simulations of active particles~\cite{paul2023dynamical}. These correlations are much harder to measure experimentally, despite their fundamental relevance to reveal the underlying microscopic processes relevant to emerging glassiness.   

\begin{figure}
\begin{center}
\includegraphics[width = \linewidth]{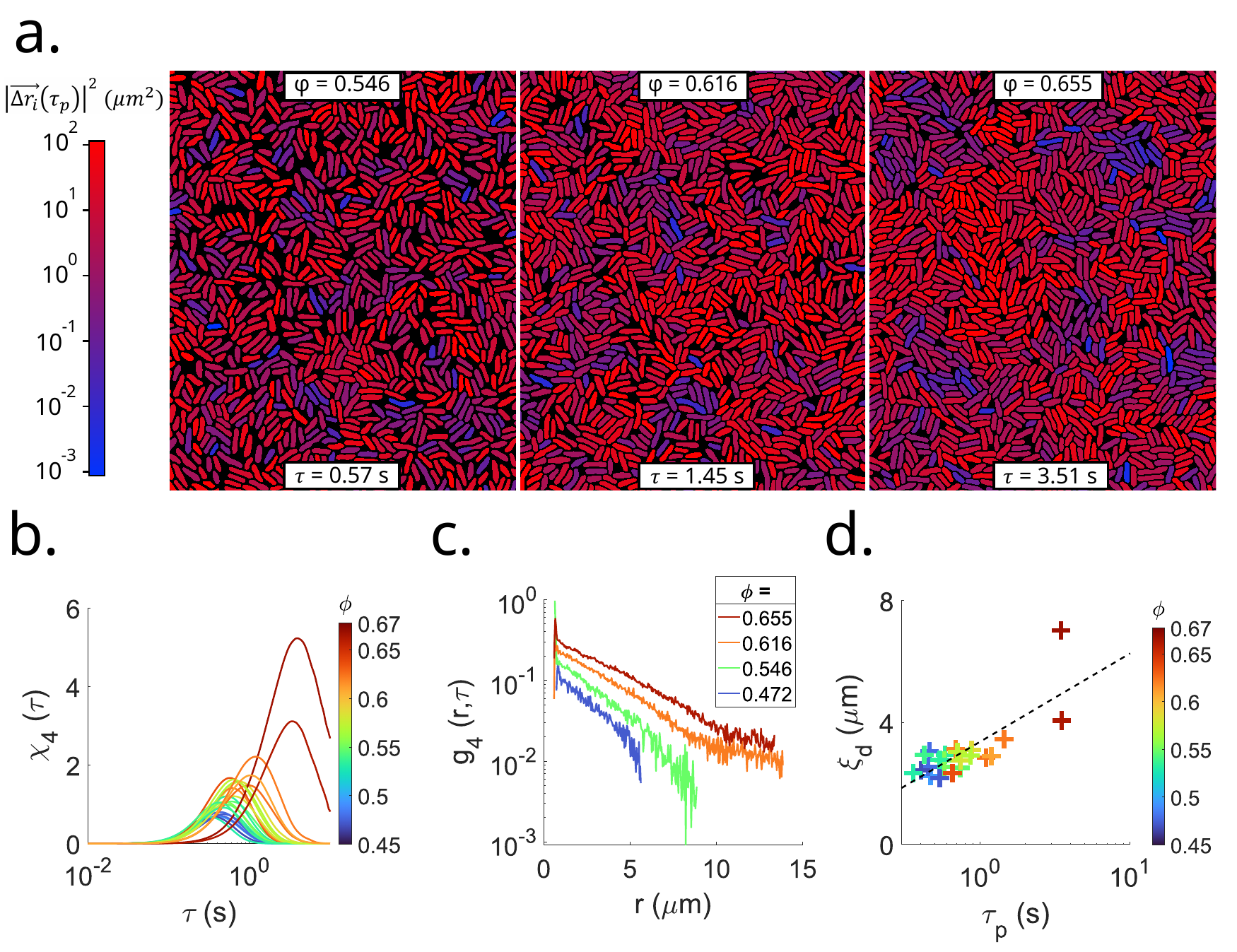}\\
\caption{\textbf{Growth of spatially heterogeneous glassy dynamics.} 
(\textbf{A}) The maps of squared displacements at time $\tau_p$ for three different surface fractions reveal growing spatial correlations between mobile (red) and immobile (blue) regions. 
(\textbf{B}) Time dependence of the four-point dynamic susceptibility $\chi_4(\tau)$ for different surface fractions. 
(\textbf{C}) Four-point spatial correlation functions display a nearly exponential decay over a growing dynamic length scale $\xi_d(\phi)$.  
(\textbf{D}) The parametric evolution of the dynamic correlation length $\xi_d$ with the relaxation time $\tau_p$ is well described by a logarithmic dependence, $\xi_d \propto \log \tau_p$, as indicated by the dashed line. 
Panel (\textbf{A}) only about 30\% of the full field of view, but panels (\textbf{B}-\textbf{D}) were calculated using the complete field of view.}
\label{fig:Spatial_heterogeneities}
\end{center}
\end{figure}

We start with qualitative evidence of growing spatial correlations in Fig.~\ref{fig:Spatial_heterogeneities}(A), where we color-code the amplitude of particle displacements over a duration comparable to the relaxation time $\tau_p$ (Movies~\ref{simov:WT_low_SD}, \ref{simov:WT_mid_SD}, and \ref{simov:WT_high_SD}). These maps reveal regions of high and low displacement, with a characteristic size that seems to increase with $\phi$. Direct comparisons of translational and orientational heterogeneity maps reveal a strong correlation (fig.~\ref{sifig:Relaxation}), demonstrating that fast moving bacteria also rotate fast (the corresponding statistical analysis is provided in Supplementary Text). Additionally, we gathered evidence (fig.~\ref{sifig:NoSpatialStructure}) that the growing dynamic domains are uncorrelated with several structural features, such as local surface fraction, cell size, nematic order, and tetratic order. We also confirmed that flagellar activity is maintained when cells are caged (Movie~\ref{simov:FliC}). Together, these observations demonstrate that the emerging spatial dynamic correlations are controlled by the competition between crowding and activity, and represent a novel collective feature characterizing the dynamics of dense bacterial assemblies.

Following work on equilibrium systems~\cite{berthier2011dynamical}, we quantify these growing correlations using multi-point correlation functions and susceptibilities. We first calculate the four-point dynamic susceptibility $\chi_4(\tau) = N [ \langle Q(\tau)^2 \rangle - \langle Q(\tau) \rangle^2 ]$ with $Q(\tau)=1/N \sum_i q_i(\tau)$ where the local overlap $q_i(\tau) = \exp(- |\Delta \vec{r}_i(\tau)|^2/a^2)$ (with $a=3.5\,\mu{}m$) is a convenient local indicator of motion~\cite{dauchot2005dynamical}. As shown in Fig.~\ref{fig:Spatial_heterogeneities}(B), the four-point susceptibilities peak at a time scale that increases with surface fraction and essentially tracks the evolution of $\tau_p$. Interestingly, the peak height also grows with $\phi$, which directly reveals that the typical area of dynamically correlated regions increases as the bacterial glass transition is approached~\cite{lavcevic2003spatially,toninelli2005dynamical}.

To accurately measure the linear size of correlated domains, we calculated the spatial correlations between fluctuations of the local overlaps, $g_4(r,\tau) = \langle \sum_{ij} \delta q_i (\tau) \delta q_j (\tau) \delta( r - |\vec{r}_i-\vec{r}_j|)  \rangle$, with $\delta q_i = q_i - \langle Q \rangle$, see Fig.~\ref{fig:Spatial_heterogeneities}(C). The measured four-point spatial correlation functions decay exponentially with distance, $g_4(r,\tau_p) \propto \exp(-r/\xi_d)$, which provides a determination of a characteristic correlation length $\xi_d$. As shown in Fig.~\ref{fig:Spatial_heterogeneities}(D), $\xi_d$ increases modestly with surface fraction, as evidenced by the approximate linear relation $\xi_d \propto \log(\tau_p)$. This relation, reported before for molecular fluids~\cite{dalle2007spatial}, confirms that dynamical heterogeneities grow as the system approaches a dynamically arrested glass state. A similar analysis for rotational motion (fig.~\ref{sifig:DH_SR}) provides a comparable dynamical length scale, confirming further the strong coupling between position and orientation fluctuations. 

\section*{Discussion and outlook}

Our in-depth experimental analysis reveals emerging glassy behavior across bacterial monolayers, showing a general slowdown in dynamics and growing dynamic fluctuations in space and time as density approaches the bacterial glass transition that lead to arrested solid states. Leveraging high-resolution tracking of cell motion across several orders of magnitude in both space and time, we systematically quantify these effects with remarkable precision. Our main conclusion is that dense bacterial assemblies gradually solidify via a physical process exhibiting striking quantitative similarities with its counterpart in dense colloidal and molecular fluids, while also displaying unique behavior, including unusual violations of the Stokes-Einstein relation. Notably, we observe no decoupling between orientation and position dynamics across five bacterial strains, even for elongated mutants. This robust conclusion contrasts with recent studies on \textit{E.~coli} monolayers~\cite{lama2024emergence} and of colloidal ellipsoids~\cite{zheng2011glass}, and is further supported by invoking frictional interactions between cells, as demonstrated recently in studies of elongated colloids with varying roughness~\cite{liang2025glass}.

We observed that variations in swimming behaviors -- whether cells alternate directions or swim unidirectionally -- have little to no impact on the critical surface fraction at which the glass transition occurs. Similarly, other motility features, such as multi-flagellation or the absence of type-IV pili, show no effect on glassy dynamics. These findings align well with numerical studies showing that the specifics of self-propulsion do not influence active glassy dynamics~\cite{debets2022active}. However, the elongated strain exhibits a lower critical surface fraction, emphasizing the importance of cell shape over motility details in setting the transition density.

From a biological perspective, our findings provide new insight into how dense bacterial populations reorganize under physical constraints. It is well established that \textit{P.~aeruginosa} transitions from a motile, planktonic state to a biofilm state upon surface adhesion, embedding cells within a protective extracellular matrix~\cite{lee2018multigenerational, yan2017bow}. This process is typically attributed to biochemical regulation, but our study shows that crowding alone can induce a dramatic reduction in motility, independently of any molecular commitment to the biofilm state. Rather than becoming entirely immobilized at high density, cells retain slow, correlated motion, which may support long-timescale structural remodeling, resource redistribution, or mechanical adaptation. These results suggest that crowding-induced slow-down is not merely a byproduct of biofilm formation, but may actively contribute to the transition by priming the population for matrix secretion, adhesion, and spatial patterning. 

\clearpage

\begin{figure}
\begin{center}
\includegraphics[width = \linewidth]{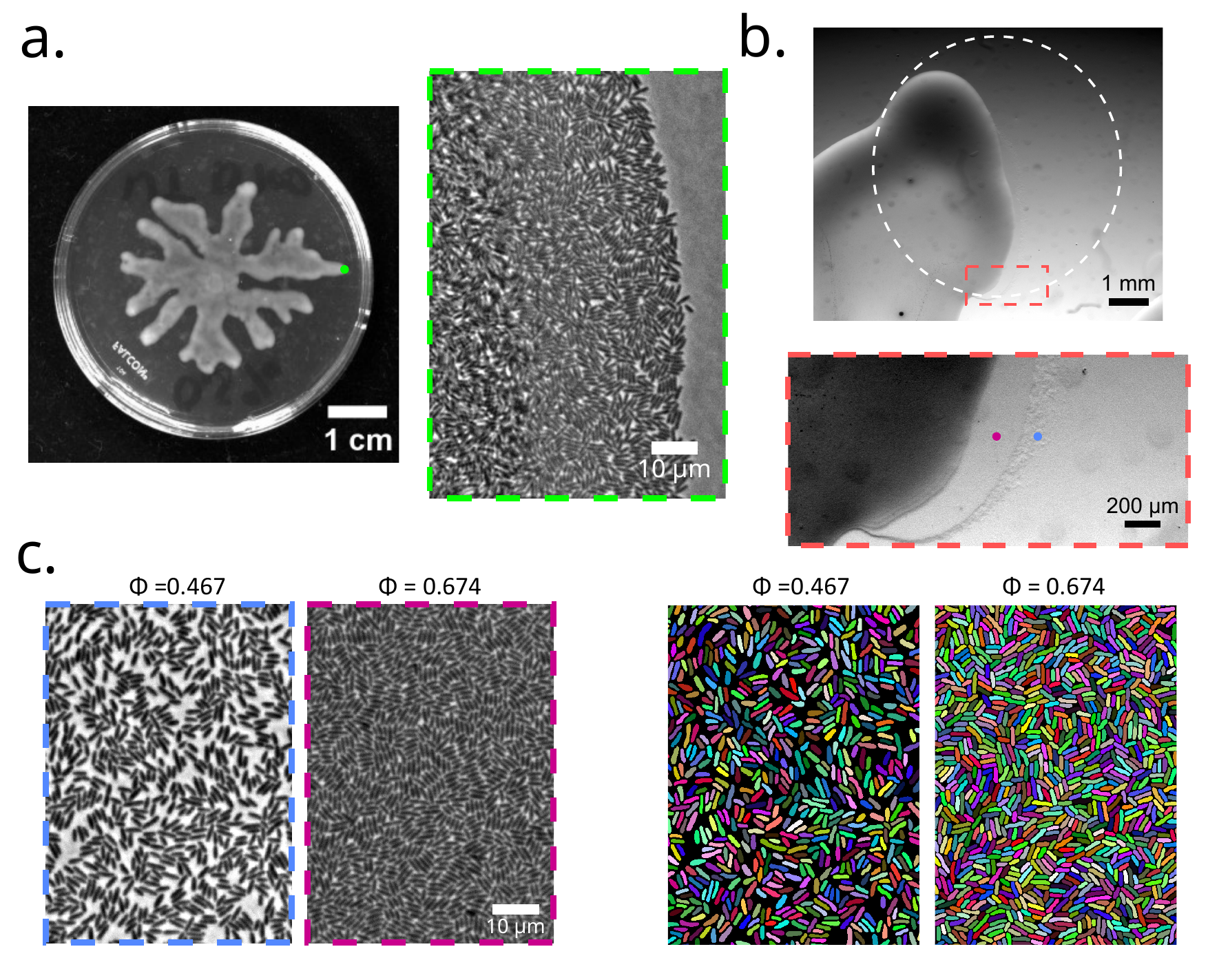}\\
\caption{\textbf{Producing and imaging large dense bacterial monolayers.} 
(\textbf{A}) Illustrative example of macroscopic colony and close-up on naturally occurring monolayer of cells at the edge. 
(\textbf{B}) Branch tip after deposition of a water droplet. The boundary of the area covered by the droplet is depicted with a white dashed line.
(\textbf{C}) Illustrative examples of pictures obtained at different surface fractions (left) and results of the segmentation and tracking performed on the images after using \textsc{Distnet2D}~\cite{ollion2024distnet2d}, with colors randomly assigned to cells.}
\label{fig:Exp_design}
\end{center}
\end{figure}

\begin{figure}
\begin{center}
\includegraphics[width = \linewidth]{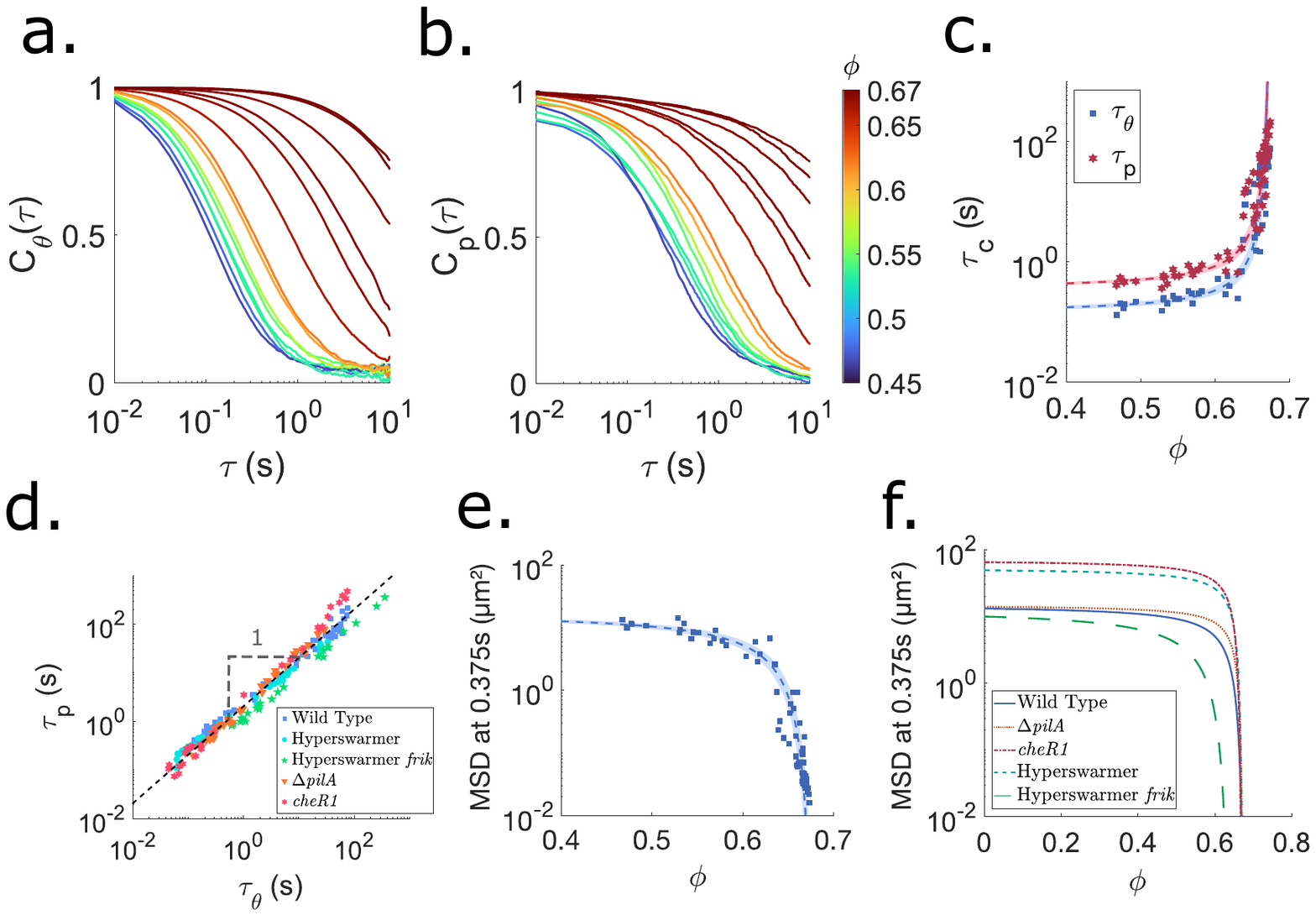}\\
\caption{\textbf{Analysis of global orientational and translational dynamics.} 
(\textbf{A}) Orientation time correlation function for different surface fractions (unless specified otherwise, all curves correspond to the wild-type \textit{P.~aeruginosa} strain). 
(\textbf{B}) Translational time correlation function for different surface fractions. 
(\textbf{C}) Evolution of correlation times for orientation and position with surface fraction. Dashed curves are fits to Eq.~(\ref{eq:VFT}). 
(\textbf{D}) Parametric plot of translational and orientational correlation times, demonstrating the linear correlation between them.
(\textbf{E}) Mean-squared displacement (MSD) at an arbitrarily chosen time $\tau=0.375$~s as a function of surface fraction. Dashed curve is a fit to  Eq.~(\ref{eq:VFT}). 
(\textbf{F}) Same as (E) compared between various mutants. For clarity, we show the individual fits to Eq.~(\ref{eq:VFT}), with raw data shown in fig.~\ref{sifig:MSD_phi_all} and fitting parameters in fig.~\ref{sifig:VFT}. 
In (C, E), shades represent standard deviation on the fitting parameters (see calculation details in the methods \cite{methods}).}
\label{fig:Bulk_dynamics}
\end{center}
\end{figure}

\newpage

\clearpage


\clearpage 

\bibliography{arxiv} 
\bibliographystyle{unsrt}

\section*{Acknowledgments}

The authors thank Lisa Sanchez for providing the PMQ30 plasmid, Susanne Häußler for the \textit{cheR1} strain, and Dominique Limoli for the $\Delta$\textit{pilA} strain. They are also grateful to Elizabeth Warren for assistance with the mutation protocol. Additional thanks go to Jean Ollion for support with training \textsc{Distnet2D}, and to Pierre Illien for fruitful discussions.
\paragraph*{Funding:}
MD and MM acknowledge the support of the French Agence Nationale de la Recherche (ANR), under grant ANR-21-CE30-0025 (project X-BACAMAT). LB acknowledges the support of the French Agence Nationale de la Recherche (ANR), under grant ANR-20-CE30-0031 (project THEMA).
\paragraph*{Author contributions: Conceptualization:} MM, LB, MD; Formal analysis: MM, MD; Funding Acquisition: LB, MD; Investigation: MM; Methodology: MM, MD, LB; Project administration: MD; Resources: MM, NFB; Software: MM; Supervision: LB, MD; Visualization: MM, MD, LB; Writing – original draft: MM, NFB, MD; Writing – review \& editing: MM, LB, MD.
\paragraph*{Competing interests: }
There are no competing interests to declare.
\paragraph*{Data and materials availability: } Data is available on a Dryad repository.


\newpage


\renewcommand{\thefigure}{S\arabic{figure}}
\renewcommand{\thetable}{S\arabic{table}}
\renewcommand{\theequation}{S\arabic{equation}}
\setcounter{figure}{0}
\setcounter{table}{0}
\setcounter{equation}{0}

\newpage


\subsection*{Materials and Methods}
\subsubsection*{Construction of \textit{fliC}$^{\text{T394C}}$ and $\Delta$\textit{cheR1} \textit{fliC}$^{\text{T394C}}$ strains}
\label{sec:Mutants_obtention}
In order to visualize flagella in \textit{P.~aeruginosa}, we introduced a T394C substitution in the flagellum protein FliC. The primers were designed according to the method previously explained \cite{de2017high}. To generate the mutant FliC(T394C) construct, DNA fragments upstream and downstream of the \textit{fliC} gene were amplified by PCR from PA14 genomic DNA using the following primer pairs FliC-up-for, T394C-up-rev, T394C-dn-for, FliC-dn-rev (sequences are listed in Table \ref{si_table:Primers}).

The $\Delta$\textit{cheR1} strains were obtained by removing the entire coding sequence of the \textit{cheR1} gene.  DNA fragments upstream and downstream of the cheR1 gene were PCR amplified from PA14 genomic DNA using the primer pairs CheR1-up-for, CheR1-up-rev, CheR1-dn-for, CheR1-dn-rev (sequences are listed in Table \ref{si_table:Primers}).

PCR products for the construction of both \textit{fliC}$^{\text{T394C}}$ and $\Delta$\textit{cheR1} strains were then cloned into pMQ30 via in vitro homologous recombination using NEBuilder HiFi DNA Assembly kits (New England Biolabs). The resulting pMQ30-FliCT394C and pMQ30-UpDw-cheR1 plasmids were used to transform \textit{E. coli} TG1 via electroporation. Plasmids were extracted and used to transform \textit{E. coli} S17 via heat shock before being subsequently introduced into \textit{P.~aeruginosa} via conjugation. Integrants were selected on a Vogel-Bonner Minimal Medium (VBMM) agar, which contains (per liter): 2 g Magnesium Sulfate Heptahydrate (MgSO$_4$-7H$_2$O), 20 g Citric Acid (C$_6$HO$_8$O$_7$), 35 g Sodium Ammonium Hydrogen Phosphate Tetrahydrate (NaNH$_4$HPO$_4$-4H$_2$O), 100 g Dipotassium Phosphate (K$_2$HPO$_4$). After sterilization by autoclaving, the medium was supplemented with 50 $\mu$g/mL gentamicin, 0.1\% Casamino Acids and 1 mM MgSO$_4$. To evict the plasmid, cells were grown in 5 mL lysogeny broth (LB) for 12-16 hours at 30$^{\circ}$C, serially diluted with LB, spread on LB agar containing 7\% sucrose, and then incubated at 30$^{\circ}$C overnight. Individual colonies were patched on LB plates and LB plates containing gentamicin 50 $\mu$g/mL to identify gentamicin-sensitive colonies that had evicted the plasmid. Colonies that had excised the plasmid were screened by PCR using the primer pairs T394C-check-for and FliC-check-rev, or the primer pairs CheR1-check-for and CheR1-check-rev for analyzing the \textit{fliC}$^{\text{T394C}}$ or $\Delta$\textit{cheR1} strains respectively. PCR products were sequenced using the primer T394C-check-for or cheR1-check-for to determine which isolates had retained either the allele encoding \textit{fliC}$^{\text{T394C}}$ or the $\Delta$\textit{cheR1} gene, respectively. 

\subsubsection*{Bacterial colonies}

Bacterial cells were grown overnight in LB medium at 37$^{\circ}$C with aeration. Agar plates of swarming medium (47 mM Na$_2$HPO$_4$, 22 mM KH$_2$PO$_4$, 8.5 mM NaCl, 1 mM MgSO$_4$, 0.1 mM CaCl$_2$, 5 g/L casamino acids (Bacto, BD)) were prepared via addition of agar to reach a 0.5\% mass fraction \cite{deforet2023long}. Overnight bacterial suspension was washed twice in phosphate-buffered saline buffer (PBS) and diluted 1000-fold. \text{2~$\mu$L} of the washed suspension were used to inoculate an agar plate, which was then flipped and placed inside a 37$^{\circ}$C microbiological incubator overnight.

\subsubsection*{Bacterial monolayers}

\label{sec:Monolayers}

Monolayers were naturally present at the edge of the swarming colonies, but they were only a few dozens of cells wide and they were intrinsically anisotropic: edge cells were nearly immotile and inner cells exchanged with the bulk (Fig.~\ref{fig:Exp_design}(A)). To obtain extended, isotropic monolayers, a $5~\mu$L water droplet was added \textit{in situ} at the edge of the colony branch to disperse cells. As the water is absorbed into the gel, cells get immobilized on the gel surface (Fig.~\ref{fig:Exp_design}(B)). Cells from the bulk of the colony then migrated to the newly colonized zone, progressively increasing the surface fraction over a period of minutes.

\subsubsection*{Imaging}

Swarming plates were placed inside an stage-top incubator (Okolab), regulated at 37$^{\circ}$C and mounted on an inverted microscope (IX-81, Olympus). Phase contrast videos of 10 seconds at 100 frames per seconds at 40$\times$ magnification were acquired using a Blackfly S camera (FLIR), with a resolution of 0.088 $\mu$m/pixel. The field-of-view was cropped from $2448\times2048$ to $1000\times1000$ pixels to reach a frame rate of 100 frames-per-second.

\subsubsection*{Image analysis}

Cell segmentation and tracking were performed using \textsc{Distnet2D}, a method described in \cite{ollion2024distnet2d}. The deep-learning model was trained on several high-density movies and on one low-density movie. Data analysis was performed with MATLAB (The MathWorks, Inc.).

\subsubsection*{Flagella labeling and imaging}

To obtain movies of rotating flagella of \textit{P.~aeruginosa} in swarming state, we used the \textit{fliC}$^{\text{T394C}}$ mutants detailed in Table \ref{si_table:Strains}. 5 $\mu$L of Alexa-568 maleimide (200$\mu$M) was added \textit{in situ} in a swarming plate assay to stain the flagella. To obtain better imaging contrast, the stained cells were then scraped out of the gel using a 1 $\mu$L loop and resuspended in a PBS buffer with 1\% mass fraction of Triton to avoid clumping in the suspension. A droplet of this suspension was then added to the tip of another branch of the swarming colony, following the protocol discussed above to obtain bacterial monolayers.

\subsubsection*{Data analysis}

\paragraph{Removal of edge cells}

\begin{figure}[!ht]
\begin{center}
\includegraphics[width = \linewidth]{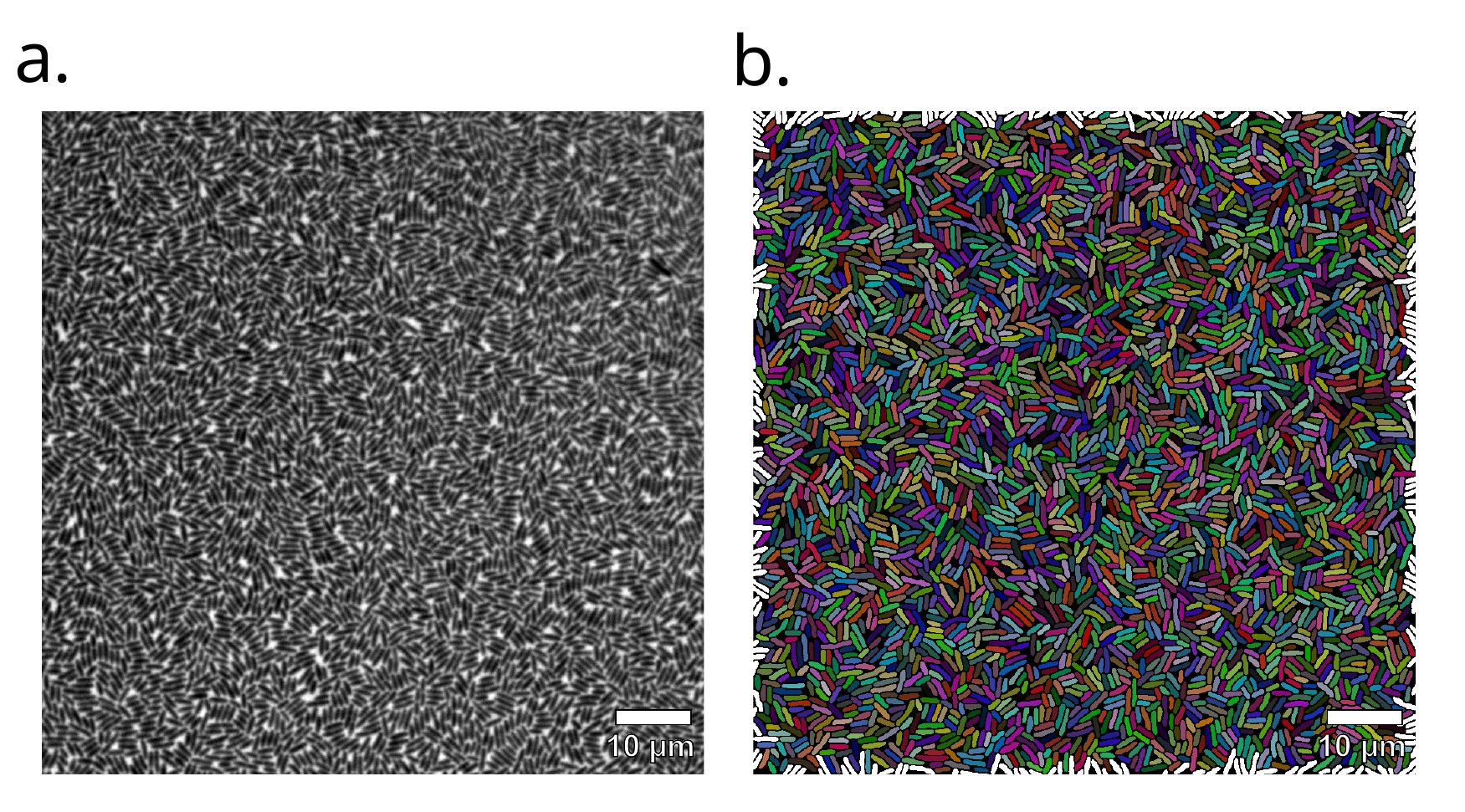}\\
\caption{\textbf{Illustration of the full field of view.} (\textbf{A}) $1000 \times 1000$ pixels image ($88 \times 88~\mu$m). For clarity, only portions of the full field of view are displayed in other figures; however, all calculations were performed using the entire field of view. (\textbf{B}) This illustrative image shows 3654 fully detected cells (random colors), while 199 cells were only partially detected cells (white) and were consequently excluded from the analysis. $\phi$ = 0.655. }
\label{sifig:all_cells}
\end{center}
\end{figure}

The segmentation and tracking algorithm is able to remarkably well detect cells at the edge of the image. Nevertheless, these edge cells are incomplete, which can make the detection of their center-of-mass hazardous and amplify small displacements, or even create some non-existent displacements. We thus decided to exclude the edge cells in all our analysis (see fig. \ref{sifig:all_cells}).

\paragraph{Removal of background motion}

As the colonies were grown on soft agar, the imaging system is very sensitive to vibrations despite the use of an optical table. This results in visible vibrations of the gel on obtained movies (see supplementary movies). To account for this, we computed the average motion of all cells from one frame to another. We then subtracted this average motion from the displacement matrix of all cells and performed all computations using these corrected displacements.

\paragraph{MSD and $\tau$ as a function of surface fraction} \label{sec:Ablation_exp}

To check to relevance of our fits on the MSD($\phi$) and $\tau$($\phi$) curves, we performed data ablation experiment: on each curve, 20\% of the data points were randomly removed, and a fit of the function was performed. This operation was repeated 1000 times, allowing to extract the mean and the standard deviation for each fitting parameter. The dashed lines on Fig. \ref{fig:Bulk_dynamics}(C) and (E) were plotted using the mean of each fitting parameter. The curve envelopes were plotted using the mean of fitting parameters $\pm$ standard deviation.

\paragraph{Diffusion coefficient computation}

To determine the infinite-time diffusion coefficient of the cells, we plotted ${MSD(\tau)} / {\tau}$ and observed whether it converged. If it did, then a diffusion coefficient was extracted as the limit of this function.

\begin{figure}[!ht]
\begin{center}
\includegraphics[width = \linewidth]{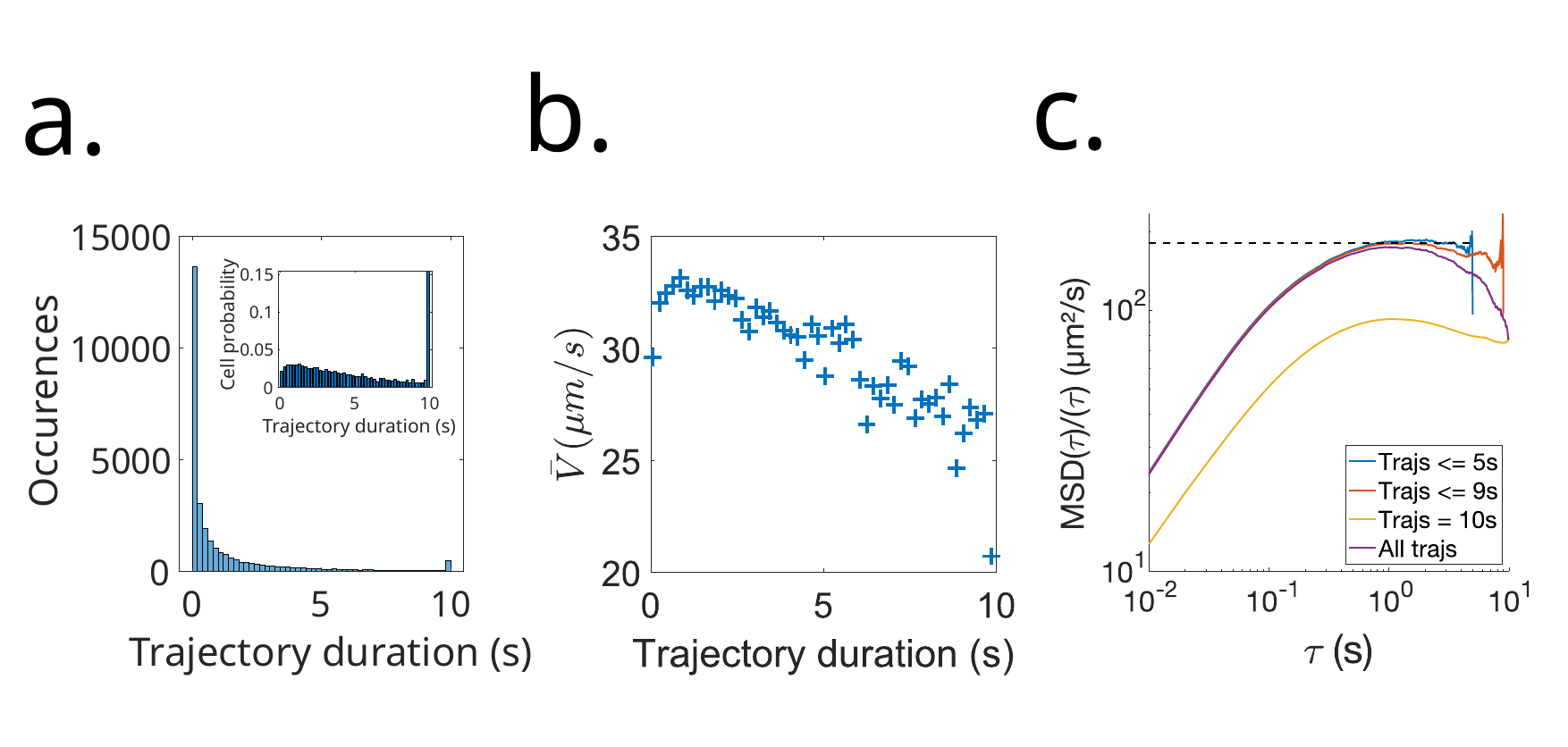}\\
\caption{\textbf{Estimation of the diffusion coefficients.} (\textbf{A}) Trajectory duration distribution, the movie is 10s long. Inset: For a given frame in the movie, probability distribution of the trajectory duration of a randomly chosen cell. (\textbf{B}) Average velocity of cells as a function of their trajectory duration. (\textbf{C}) $MSD(\tau)/\tau$ for three sub-populations, and for all trajectories, for \textit{cheR1} at low density ($\phi$ = 0.497). The black dashed line represents the limit of the curve for the trajectories shorter than 5~s, and thus this is the extracted value for the diffusion coefficient.}
\label{sifig:MSD_tau}
\end{center}
\end{figure}

We encountered an experimental limitation due to the restricted field of view. Even though our segmentation and tracking methods performed with virtually perfect accuracy, trajectories were interrupted when cells exited the field of view. Consequently, we observed a wide distribution of trajectory lengths (fig. \ref{sifig:MSD_tau}(A)), with fast-moving cells more likely to escape as represented in fig. \ref{sifig:MSD_tau}(B): short trajectories corresponded to fast cells. Thus, as time delay increased, the statistical weight of longer trajectories, \textit{ie} slower cells, increased. The MSD at a low surface fraction ($\phi$ = 0.497) \textit{cheR1} movie, computed on all trajectories, on complete (10s long) trajectories only, on trajectories shorter than 9s (excluding longer trajectories) and on trajectories shorter than 5s are presented in fig. \ref{sifig:MSD_tau}(C). At short times, the MSD on all trajectories coincided with the ones computed on short trajectories, as the MSD was dominated by the fast cells. As time increased, fast cells left the field of view and the weight of slower cells increased: the MSD computed on all trajectories caught up with the MSD computed on complete trajectories only. As the MSD, the diffusion coefficient was also dominated by fast cells, therefore for an accurate computation of the diffusion coefficient, it was necessary to look at the shorter trajectories (\textit{ie} faster cells). However, for a system approaching the glass transition, the diffusive regime was not always reached within the considered cutoff time. Thus, the shortest possible cutoff time for which the ${MSD(\tau)}/{\tau}$ curve converged was chosen (in practice, one of the three mentioned sub-populations was selected). In the example of fig. \ref{sifig:MSD_tau}(C), we selected the trajectories shorter than 5~s to calculate the diffusion coefficient.

\paragraph{Computation of $\chi_4(\tau)$ for translation}
\label{sec:Chi_4_SD_computation}

$\chi_4(\tau)$ is equivalent to the volume integral of the $g_4(r,\tau)$. However, the correlation on the local overlap $q_i(\tau)$ can get very noisy after a few $\mu$m (not shown in Fig. \ref{fig:Spatial_heterogeneities}(C)). Computing $\chi_4$ on the whole field of view can thus lead to a very noisy measurement. To account for this, while still preserving the high statistics of our measurements, we chose to cut the long distance computation of the $\chi_4$: for each $t_0$, the field of view (originally $88 \times 88~\mu$m) was divided into 25 sub-windows of $17.6 \times 17.6~\mu$m. $\chi_4(\tau)$ was independently computed considering all cells initially present in each sub-window, and then averaged over all sub-windows. This allowed all short and medium distance pairs to be considered, while ignoring the long distance pairs, whose signal-to-noise ratio was insufficient.

\paragraph{Computation of $\chi_4(\tau)$ and $g_4(r,\tau)$ for rotation}
\label{sec:Chi_4_SR_computation}
We performed rotation calculations in a manner similar to those for translation. We first calculate the four-point dynamic susceptibility $\chi_4^\theta(\tau) = N [ \langle Q_\theta(\tau)^2 \rangle - \langle Q_\theta(\tau) \rangle^2 ]$ with $Q_\theta(\tau)=1/N \sum_i q_i^\theta(\tau)$ where the local overlap $q_i^\theta(\tau) = \exp(- |\Delta \theta_i(\tau)|^2/a_\theta^2)$ (with $a_\theta=1.75\,rad$) is a convenient local indicator of rotation. To accurately measure the linear size of correlated domains, we calculated the spatial correlations between fluctuations of the local overlaps, $g_4^\theta(r,\tau) = \langle \sum_{ij} \delta q_i^\theta (\tau) \delta q_j^\theta (\tau) \delta( r - |\vec{r}_i-\vec{r}_j|)  \rangle$, with $\delta q_i^\theta = q_i^\theta - \langle Q _\theta\rangle$, see Fig.~\ref{fig:Spatial_heterogeneities}(C). The measured four-point spatial correlation functions decay exponentially with distance, $g_4^\theta(r,\tau_p) \propto \exp(-r/\xi_d^\theta)$, which provides a determination of a characteristic correlation length $\xi_d^\theta$.


\subsection*{Supplementary Text}
\subsubsection*{Fisher's exact test for position and orientation coupling}

\begin{figure}[!ht]
\begin{center}
\includegraphics[width = \linewidth]{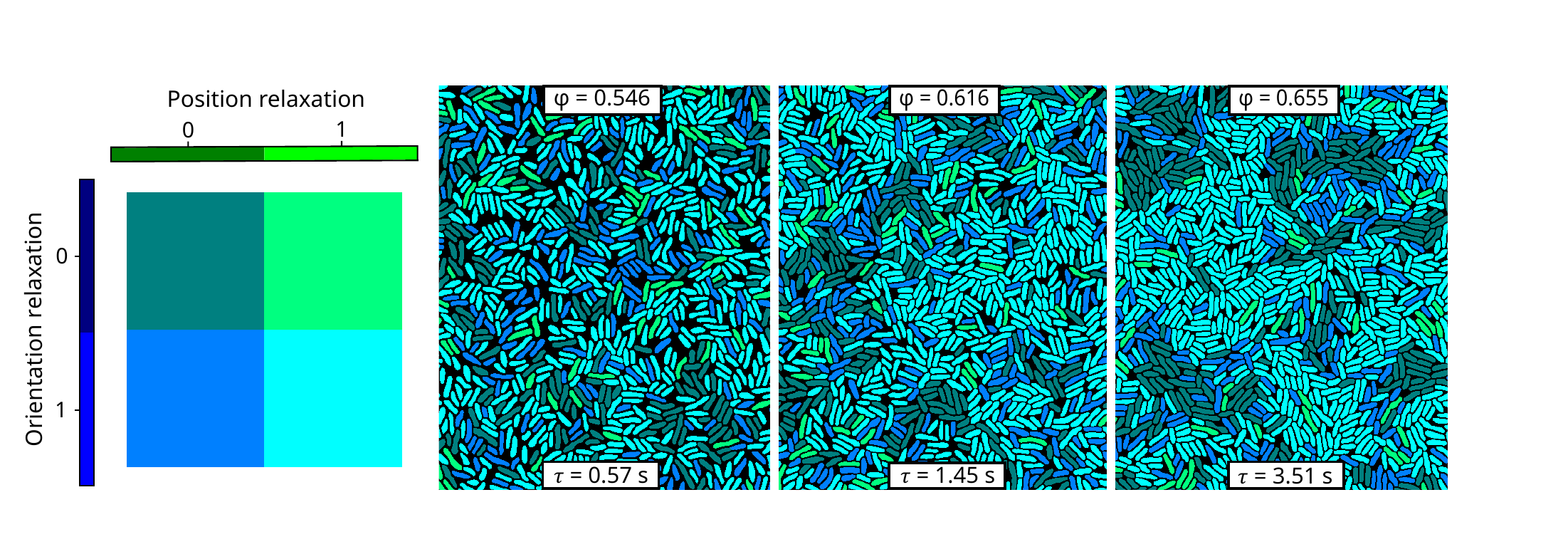}\\
\caption{\textbf{Orientation and position relaxation maps.} Orientation (in blue) and position (in green) relaxation maps for the same three snapshots as in Fig. \ref{fig:Spatial_heterogeneities}(A). Dark green represents cells that have not relaxed their orientation nor their position. Light blue represents cells that have relaxed their orientation and their position. Bright green represents cells that have relaxed their position but not their orientation. Bright blue represents cells that have relaxed their orientation but not their position.}
\label{sifig:Relaxation}
\end{center}
\end{figure}

We defined a cell as having relaxed its initial orientation when it turned by an angle of 45$^{\circ}$ or more. Similarly, we defined a as having relaxed its initial position when its center of mass moved a distance greater than the average cell body length.
To assess whether orientation relaxation and position relaxation are coupled, we studied the binary state (Position relaxed; Orientation relaxed) by constructing a contingency table. We performed a Fisher's exact test on this table for the three movies presented in Fig. \ref{fig:Spatial_heterogeneities}(A). The p-values obtained were $10^{-111}$, $10^{-137}$, and $10^{-136}$ for the three movies, indicating a strong coupling between the two relaxation modes: cells tended to turn their bodies while moving and vice versa. This result was visually supported by relaxation maps shown in fig. \ref{sifig:Relaxation}: most of the cells either relaxed both their orientation and position or neither. Few cells relaxed their position only or their relaxation only.

\newpage
\begin{figure}[!ht]
\begin{center}
\includegraphics[width = \linewidth]{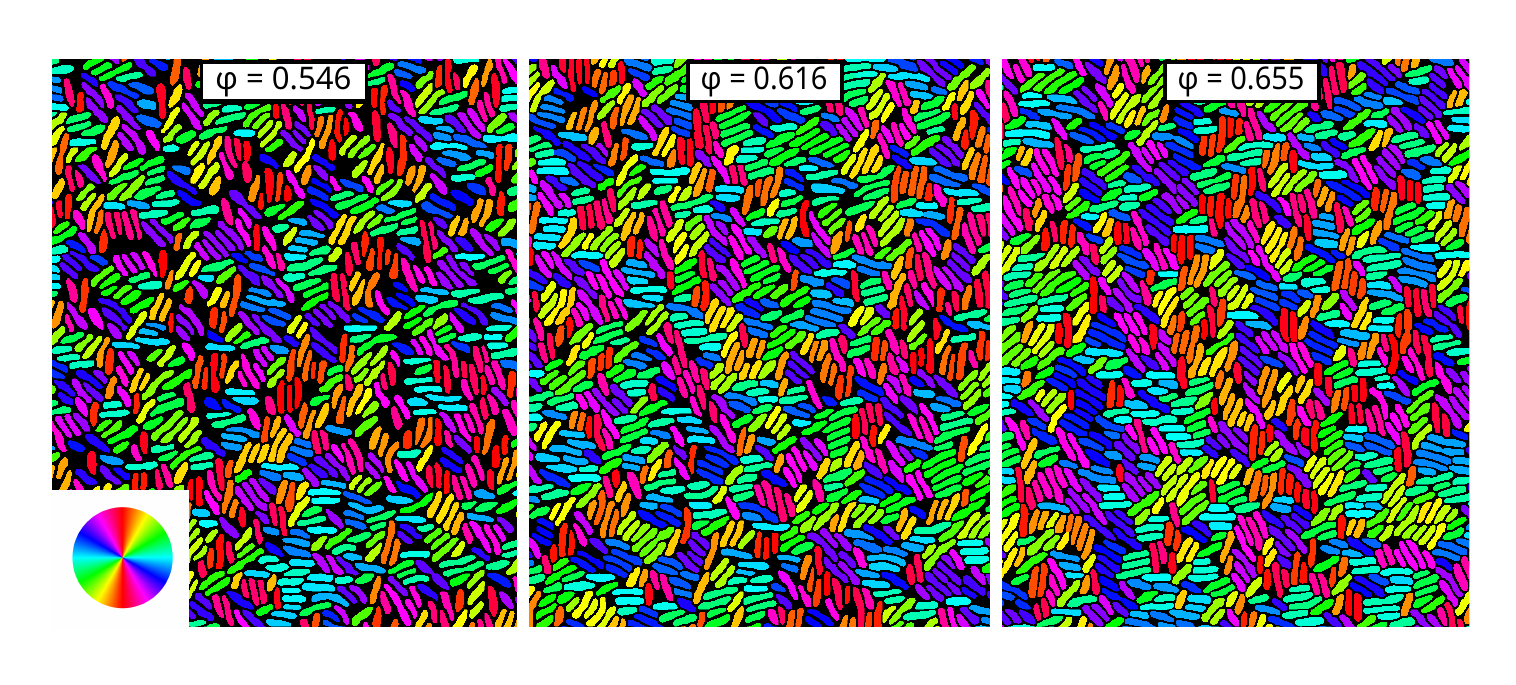}\\
\caption{\textbf{Maps of orientation do not exhibit long-range order.} Each cell is colored by the orientation of their body, computed by ellipse fitting. The color legend is $\pi$-periodic to account for cell symmetry.}
\label{sifig:Orientation}
\end{center}
\end{figure}

\begin{figure}[!ht]
\begin{center}
\includegraphics[width = \linewidth]{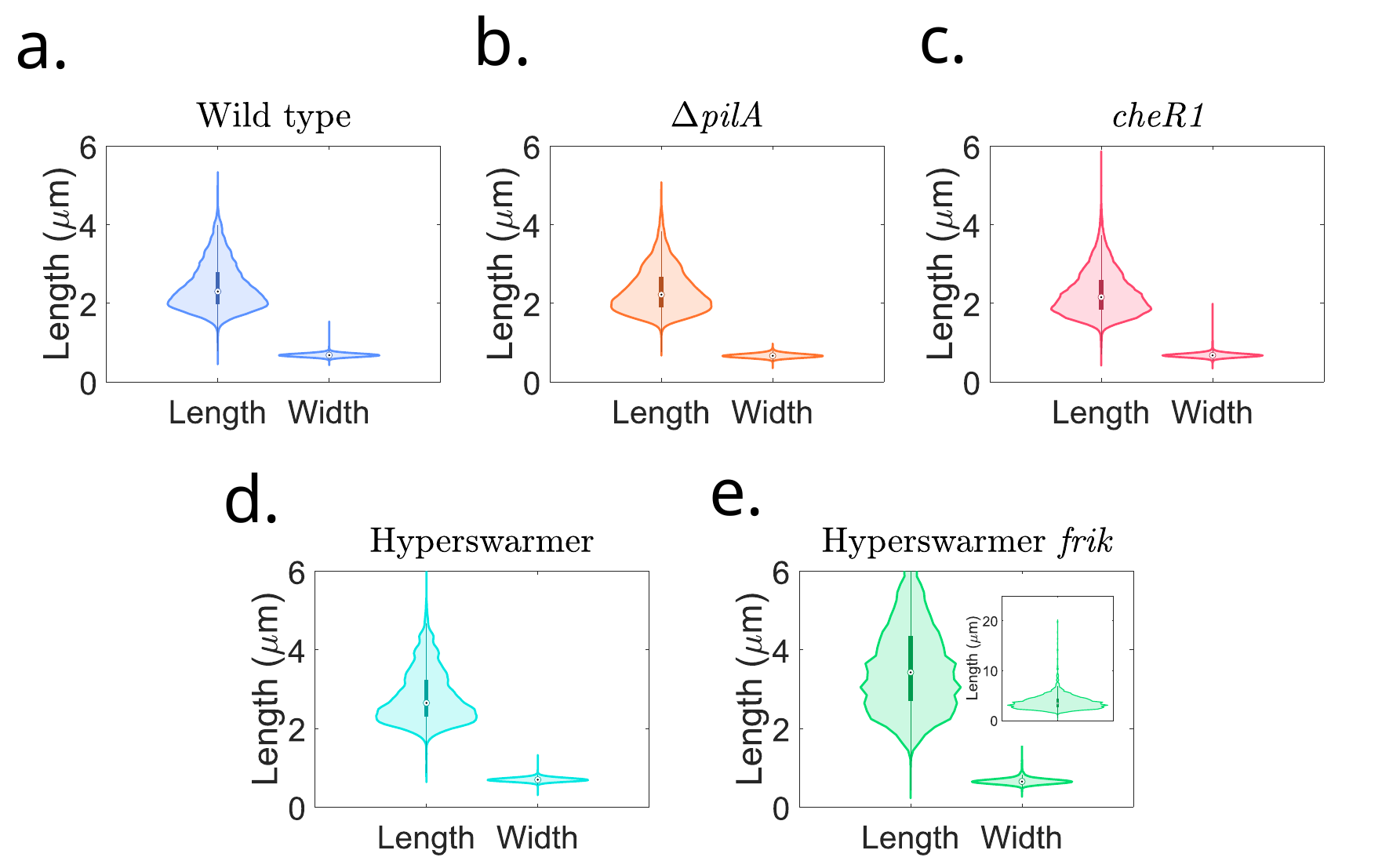}\\
\caption{\textbf{Length and width distributions for each strain.} (\textbf{A}) Wild type. (\textbf{B}) $\Delta$\textit{pilA}. (\textbf{C}) \textit{cheR1}. (\textbf{D}) Hyperswarmer. (\textbf{E}) Hyperswarmer \textit{frik}, inset: violin plot for length distribution in hyperswarmer \textit{frik} with a wider vertical range to visualize the whole length distribution.}
\label{sifig:Length}
\end{center}
\end{figure}

\begin{figure}[!ht]
\begin{center}
\includegraphics[width = \linewidth]{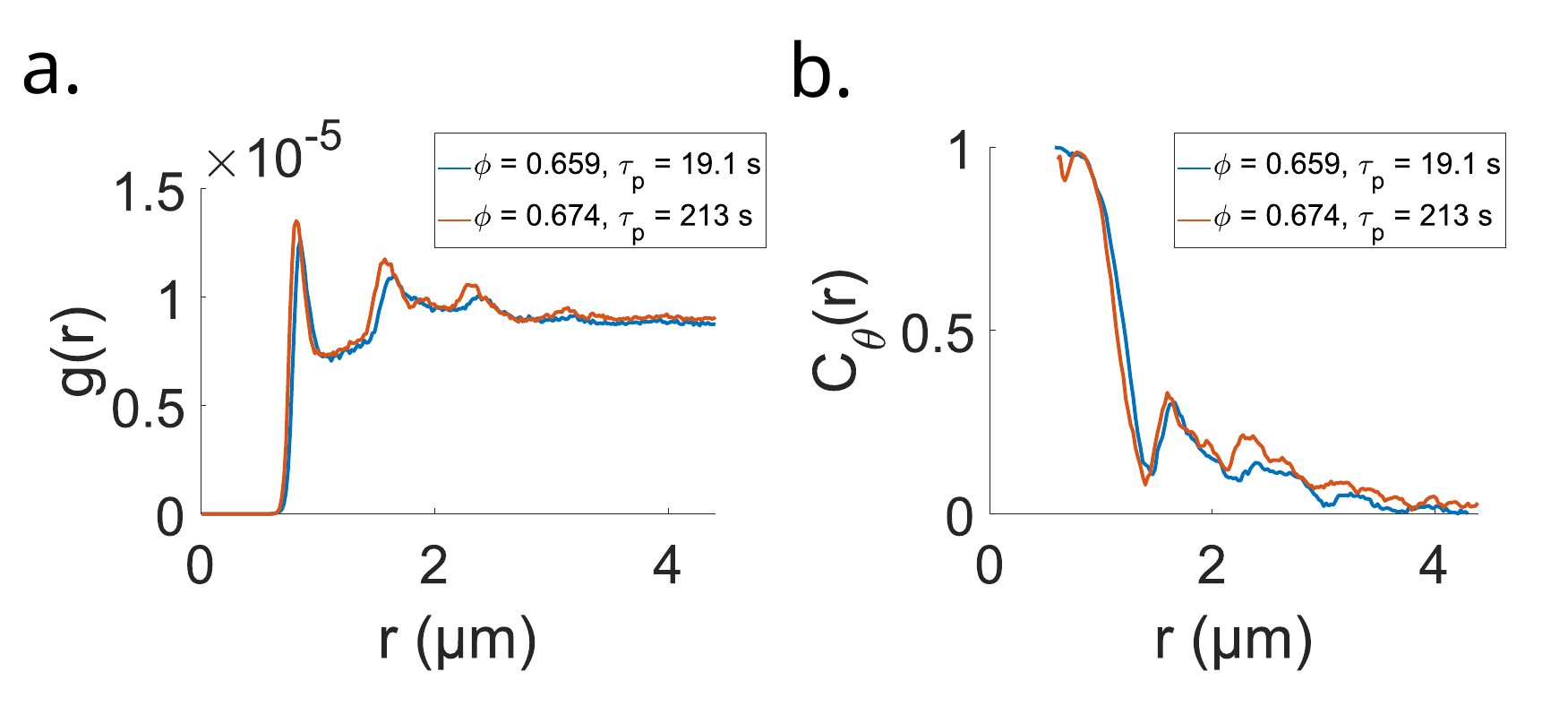}\\
\caption{\textbf{Structure is not affected by increasing surface fraction.} (\textbf{A}) Pair correlation function $g(r)$ for two surface fractions corresponding to a 10-fold increase of position correlation time. (\textbf{B}) Pair angular correlation function $C_{\theta}(r)$, defined as $C_{\theta}(r) = \langle{}\cos(2 (\theta_i-\theta_j))\rangle{}_{i,j; d_{ij} = r}$, for two surface fractions corresponding to a 10-fold increase of position correlation time.}
\label{sifig:Spatial_structure}
\end{center}
\end{figure}

\begin{figure}[!ht]
\begin{center}
\includegraphics[width = \linewidth]{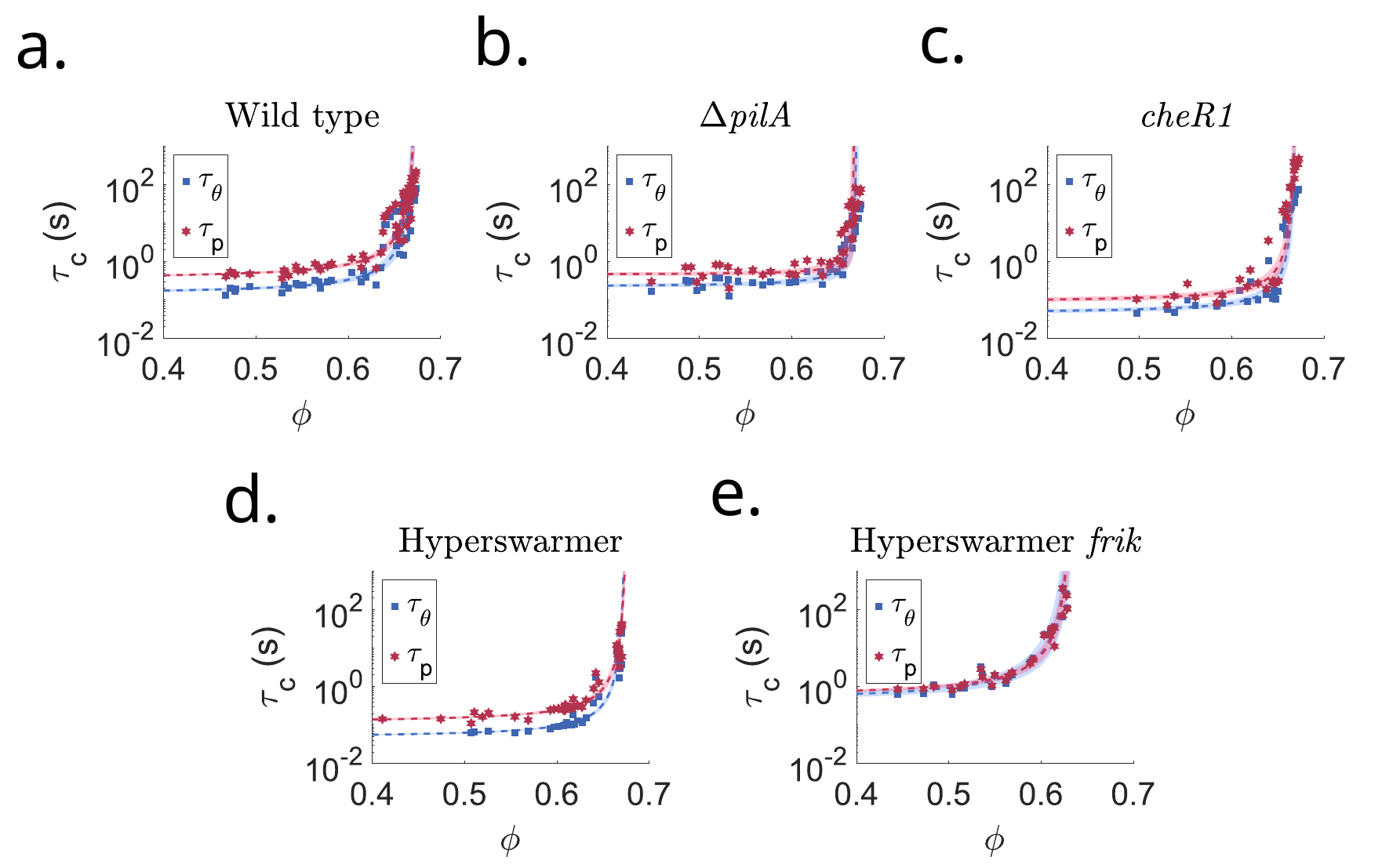}\\
\caption{\textbf{Data points and VFT fitting curves for $\tau_p(\phi)$ and $\tau_{\theta}(\phi)$ for all strains.} Each point represents one movie analyzed. The dashed line is the VFT fit obtained. The curve envelope is obtained by performing ablation experiments described in the methods \cite{methods}. (\textbf{A}) Wild type, (\textbf{B}) $\Delta$\textit{pilA}, (\textbf{C}) \textit{cheR1}, (\textbf{D}) Hyperswarmer, (\textbf{E}) Hyperswarmer \textit{frik}.}
\label{sifig:tau_phi_all}
\end{center}
\end{figure}

\begin{figure}[!ht]
\begin{center}
\includegraphics[width = \linewidth]{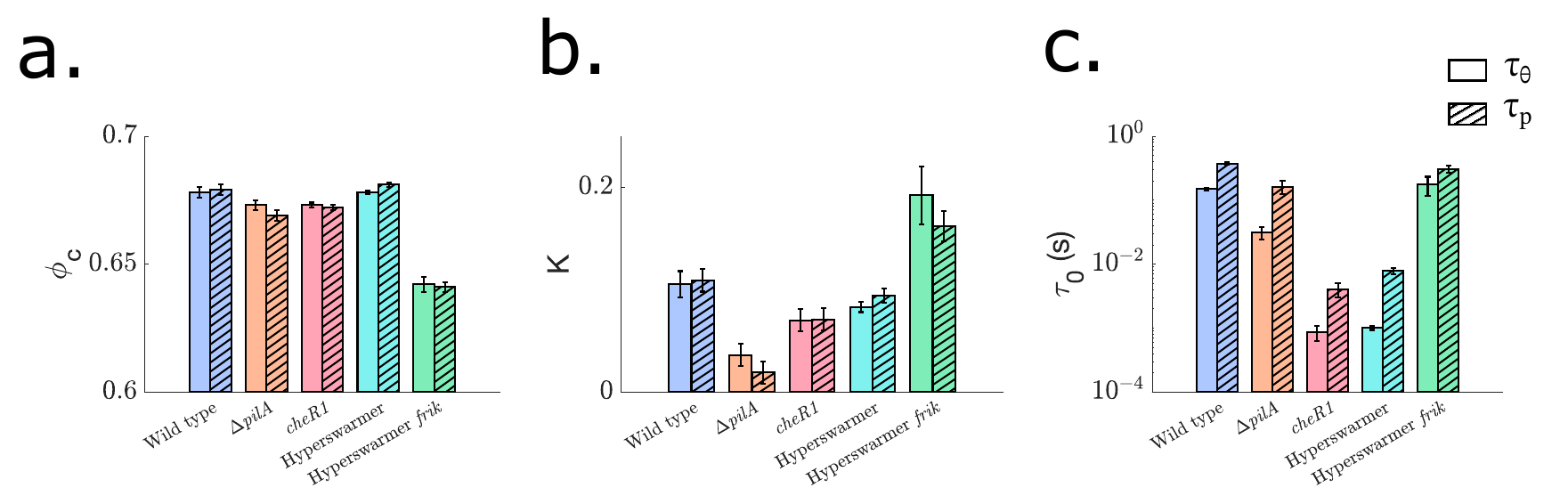}\\
\caption{\textbf{VFT fitting parameters for the curves $\tau_\theta$($\phi$) and $\tau_p$($\phi$), for all strains.} Error bars are standard deviations, obtained with the ablation fitting procedure described in the methods \cite{methods}. The plain colors represent $\tau_\theta(\phi)$, while the hatched colors represent $\tau_p(\phi)$ (\textbf{A}) Critical surface fraction $\phi_c$. (\textbf{B}) Fragility K. (\textbf{C}) $\tau_0$.}
\label{sifig:VFT_times}
\end{center}
\end{figure}

\begin{figure}[!ht]
\begin{center}
\includegraphics[width = \linewidth]{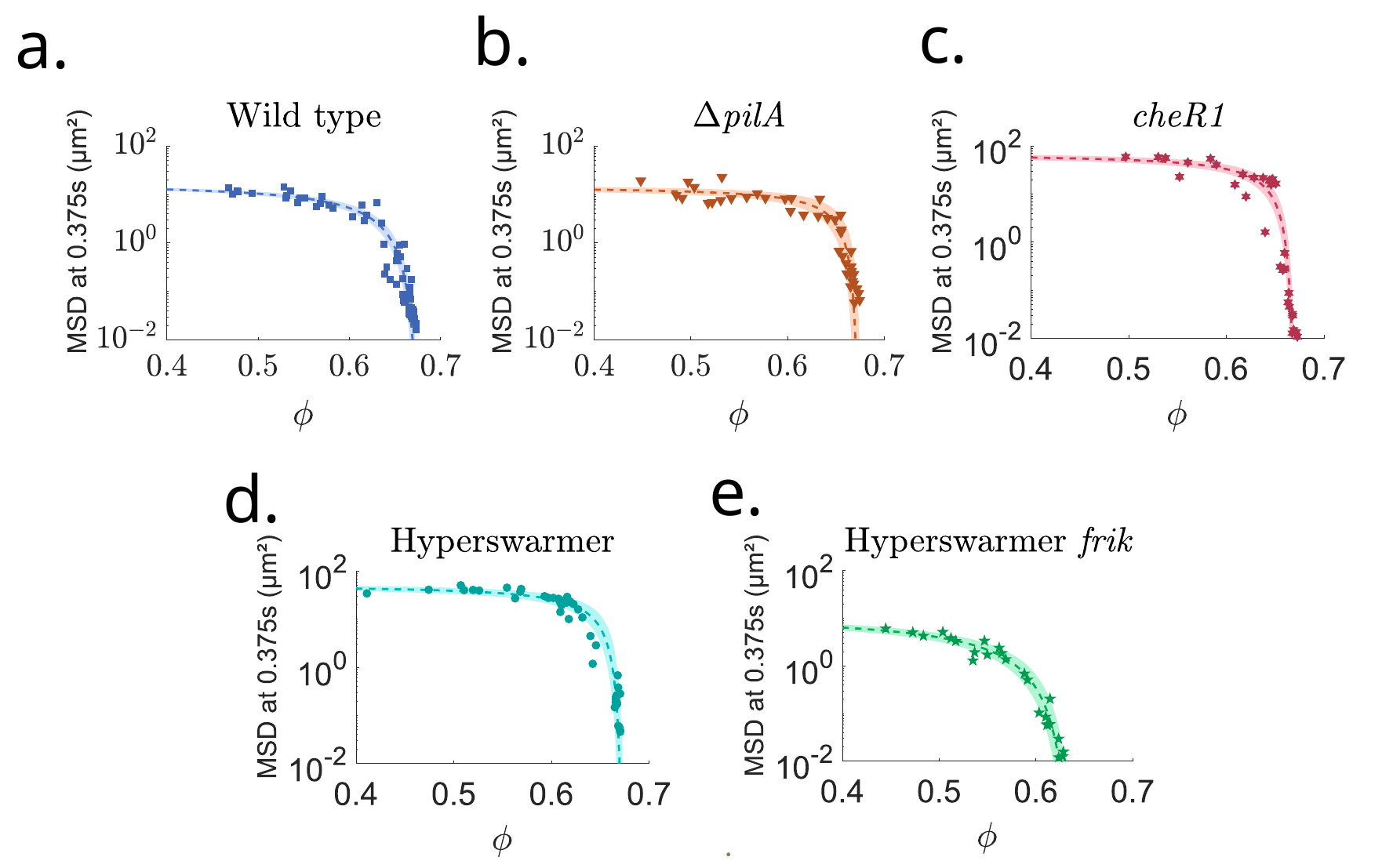}\\
\caption{\textbf{Data points and VFT fitting curves for MSD($\phi$) for all strains.} Each point represents one movie analyzed. The dashed line is the VFT fit obtained. The curve envelope is obtained by performing ablation experiments described in the methods \cite{methods}. (\textbf{A}) Wild type, (\textbf{B}) $\Delta$\textit{pilA}, (\textbf{C}) \textit{cheR1}, (\textbf{D}) Hyperswarmer, (\textbf{E}) Hyperswarmer \textit{frik}.}
\label{sifig:MSD_phi_all}
\end{center}
\end{figure}

\begin{figure}[!ht]
\begin{center}
\includegraphics[width = \linewidth]{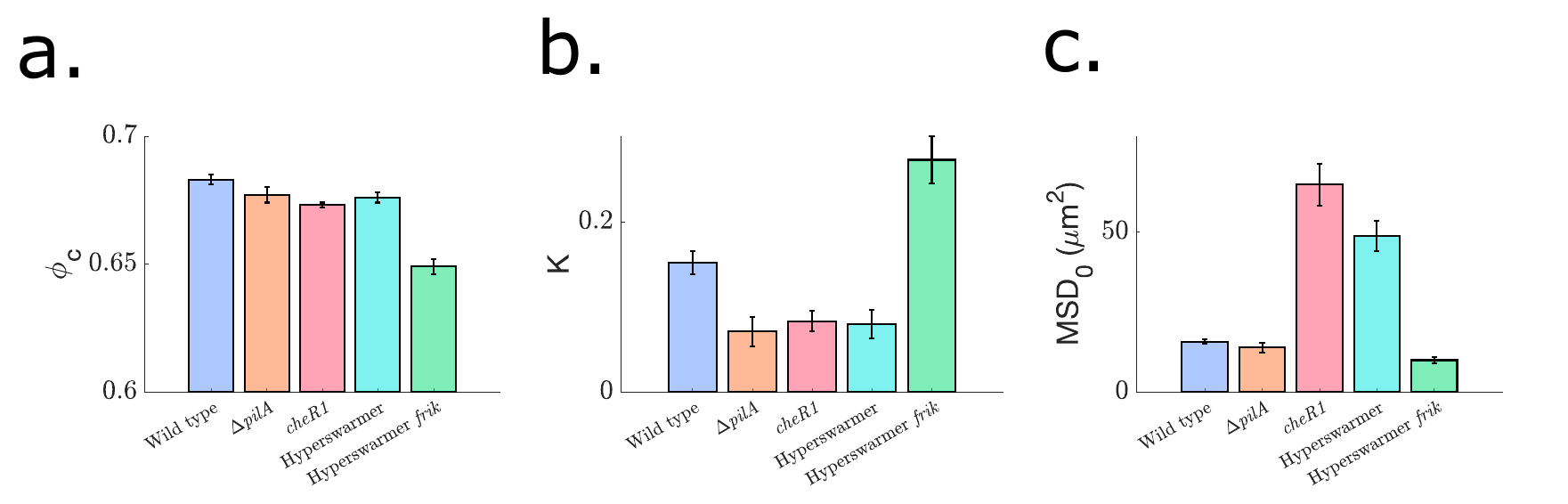}\\
\caption{\textbf{VFT fitting parameters for the curve MSD($\phi$), for all strains.} Error bars are standard deviations, obtained with the ablation fitting procedure described in the methods \cite{methods}. (\textbf{A}) Critical surface fraction $\phi_c$. (\textbf{B}) Fragility K. (\textbf{C}) MSD$_0$.}
\label{sifig:VFT}
\end{center}
\end{figure}

\begin{figure}[!ht]
\begin{center}
\includegraphics[width = \linewidth]{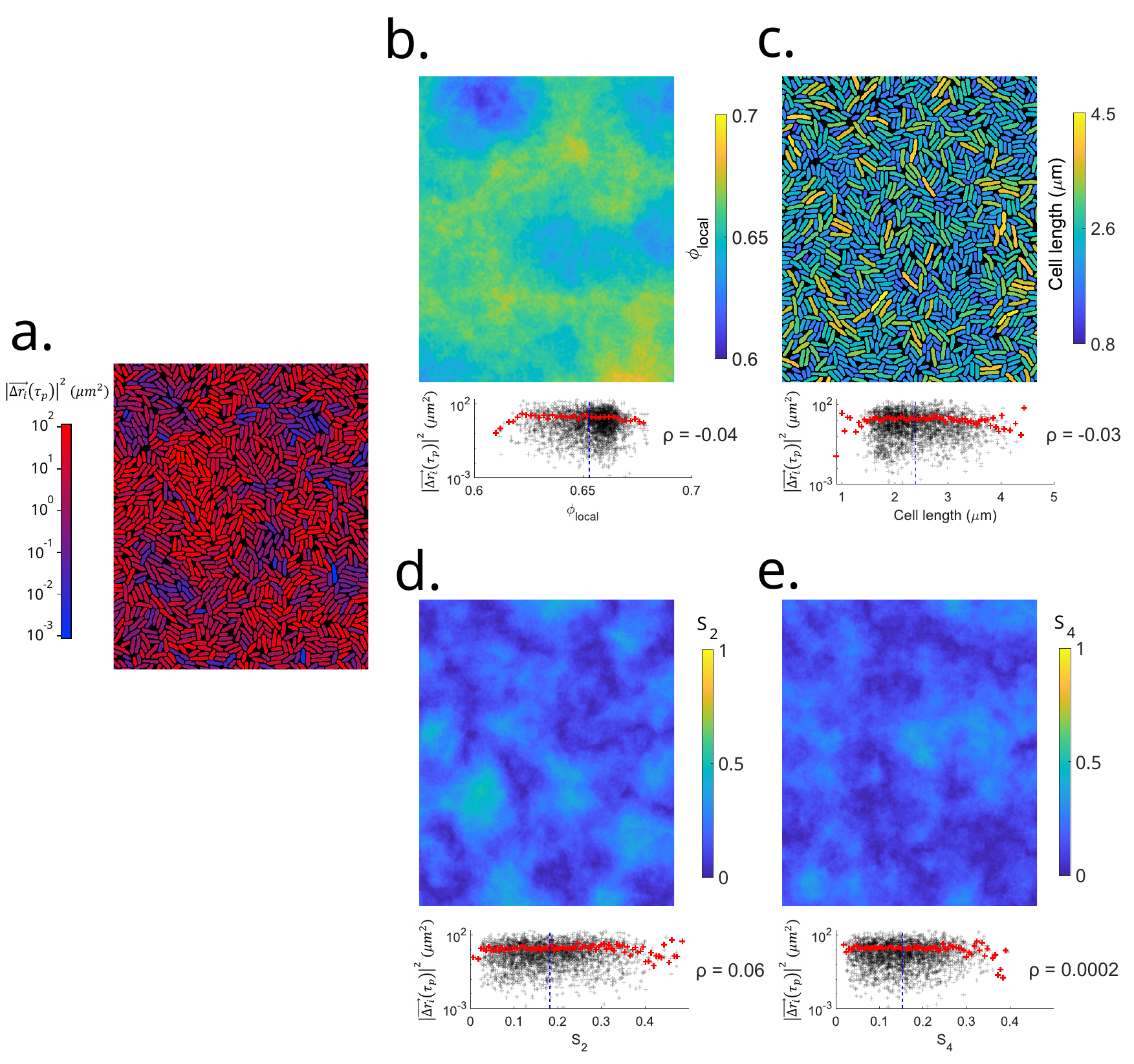}\\
\caption{\textbf{Testing for structural correlations.} (\textbf{A}) Squared displacement map for $\phi=0.655$ ($\tau_p$ = 3.51~s), reproduced from Figure \ref{fig:Spatial_heterogeneities}A. (\textbf{B}) Local surface fraction field. (\textbf{C}) Cell length map. (\textbf{D}) Local nematic order parameter ($S_2$) field. (\textbf{E}) Local tetratic order parameter ($S_4$) field. Below each panel (\textbf{B}-\textbf{E}), the squared displacement at $\tau_p$ is plotted as a function of the corresponding parameter, with gray crosses representing individual cells within the field of view and red crosses indicating bin-averaged values; the average for all cells is depicted as a blue vertical dashed line. Fields in panels (\textbf{B})-(\textbf{E}) are calculated from the snapshot as panel (\textbf{A}), using a circular kernel with a diameter of 3 cell lengths ($7~\mu$m). The Pearson correlation coefficient $\rho$ is computed for each relationship and displayed on the graph. All images represent 30\% of the full field of view to improve visualization, but the data points and the Pearson correlation coefficients were calculated using the entire field of view.}
\label{sifig:NoSpatialStructure}
\end{center}
\end{figure}

\begin{figure}[!ht]
\begin{center}
\includegraphics[width = \linewidth]{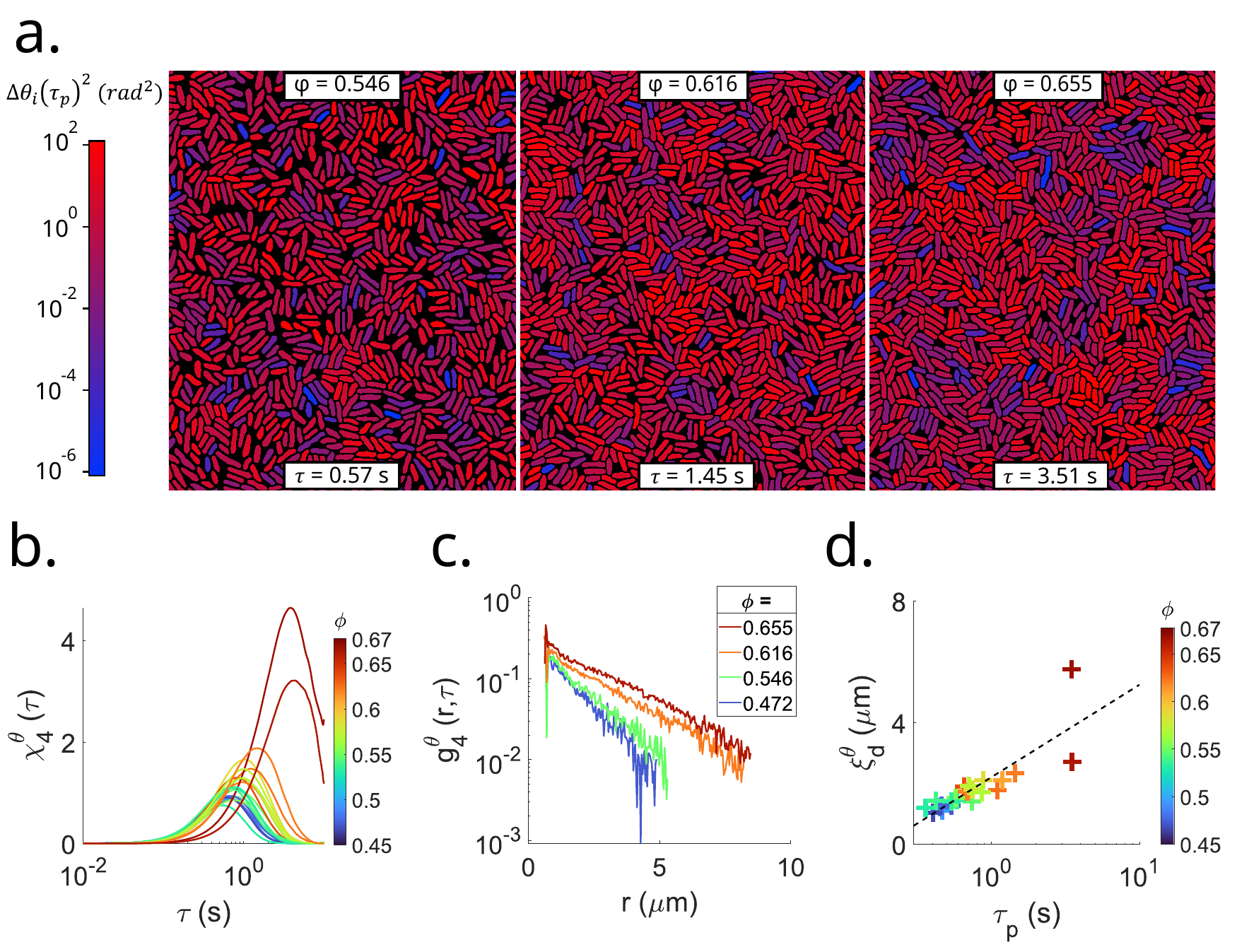}\\
\caption{\textbf{Dynamics heterogeneities on rotational motion.} (\textbf{A}) Squared Rotations (SR) maps at $\tau_p$ for three different surface fractions. (\textbf{B}) Dynamic susceptibility $\chi_4^\theta$ on rotational motion as a function of time delay for different surface fractions. (\textbf{C}) Correlation coefficient on the rotational motion at $\tau_p$ for four different surface fractions. The long-time part of the curves is not represented due to insufficient signal-to-noise ratio. (\textbf{D}) Extracted characteristic length of the rotational motion correlation function as a function of $\tau_p$ (x-axis) and surface fraction (coded in color). The dashed line represents a logarithmic fit.}
\label{sifig:DH_SR}
\end{center}
\end{figure}

\clearpage

\begin{table} 
	\caption{\textbf{List of strains used in this study and their properties.}}
\centering
\renewcommand{\arraystretch}{1.5}
\centerline{
\footnotesize
\begin{tabular}{|c|c|c|c|c|c|} 
\hline
    \textbf{Name} & \textbf{Flagellum} & \textbf{Type IV pili} & \textbf{Aspect ratio} & \textbf{Flagellar motion} & \textbf{Origin and Reference}\\ \hline
    Wild type & One & Yes & 3.89 $\pm$ 0.97 & Forward-backward & \parbox{0.3\textwidth}{\textit{Pseudomonas aeruginosa} PA14} \\ \hline
    Hyperswarmer & Multiple & Yes & 3.54 $\pm$ 0.93 & Forward-backward & \parbox{0.3\textwidth}{\vspace{0.5em}Point mutation FleN(V178G), described in \cite{deforet2014hyperswarming} (Clone 4 of \cite{van2013convergent})\vspace{0.5em}} \\ \hline
    Hyperswarmer \textit{frik} & Multiple & Yes & 5.62 $\pm$ 1.84 & Forward-backward & \parbox{0.3\textwidth}{\vspace{0.5em}Point mutation FleN(W253C), and a 9-bp deletion in \text{PA14{\_}65570}, described in \cite{deforet2015cell} (Clone 5 of \cite{van2013convergent})\vspace{0.5em}} \\  \hline
    $\Delta$\textit{pilA} & One & No & 3.90 $\pm$ 1.00 & Forward-backward & \parbox{0.3\textwidth}{\vspace{0.5em}Clean deletion of gene \textit{pilA}, \cite{Kuchma2010}, generous gift of Dominique Limoli\vspace{0.5em}}\\  \hline
    \textit{cheR1} & One & Yes & 3.26 $\pm$ 0.81 & Only forward & \parbox{0.3\textwidth}{\vspace{0.5em}\textit{cheR1} transposon mutant from the NR PA14 transposon mutant library \cite{liberati2006ordered}, described in \cite{schmidt2011pseudomonas}, generous gift of Susanne Häußler\vspace{0.5em}}\\ \hline
    Wild type \textit{fliC}$^{\text{T394C}}$ & One & Yes & 3.55 $\pm$ 0.92 & Forward-backward & \parbox{0.3\textwidth}{\vspace{0.5em}Point mutation FliC(T394C) that allows for Alexa-maleimide labeling of the flagellum, described in \cite{de2017high}. This study.\vspace{0.5em}}\\ \hline
\end{tabular}}
\label{si_table:Strains}
\end{table}

\begin{table} 
	\caption{\textbf{List of primers used in this study.} The lowercase letters indicate sequences complementary to the cloning vector pMQ30. The T394C codon mutation is underlined in the T394C-up-rev and T394C-dn-for primer sequences. The letters in italic indicate the end of the sequence from the upstream region of the \textit{cheR1} gene.}\label{si_table:Primers}
\centering
\renewcommand{\arraystretch}{1.5}
\centerline{
\footnotesize
\begin{tabular}{|l|l|} 
\hline
    \textbf{Name} & \textbf{Sequence} \\ \hline
    FliC-up-for  & tgtaaaacgacggccagtgccaagcttgcatgcctgCGACCTCAACACCTCGTTGCA  \\ 
    T394C-up-rev  & GTTCTGGGCGCCGTCGGC\underline{GCA}GGAGATGTCGACGCTGGCAACGCT \\
    T394C-dn-for  & AGCGTTGCCAGCGTCGACATCTCC\underline{TGC}GCCGACGGCGCCCAGAAC\\
    FliC-dn-rev  & ggaaacagctatgaccatgattacgaattcgagctcCGCGCTGATCGCACTCTTGA\\
    CheR1-up-for  & gcctgcaggtcgactctagaggatcGAAGAGATCCATCCGCCACC\\
    CheR1-up-rev  & GCATAAGCCTCTTCGCCCTG\\
    CheR1-dn-for  & \textit{gtcgcccagggcgaagaggcttatgc}CGCAAGGAAGCGGACCCG\\
    CheR1-dn-rev  & gtcgcccagggcgaagaggcttatgcCGCAAGGAAGCGGACCCG\\
    T394C-check-for  & cgacaagggtgtactgaccatca\\
    FliC-check-rev  & gcgctcgccttgagaatgtct\\
    CheR1-check-for  & GATGGTGAAGAAGGTCGGTG\\
    CheR1-check-rev  & CTGTCAATACAACTAGATCGCG\\ \hline
\end{tabular}
}
\end{table}


\clearpage %
\renewcommand{\figurename}{Movie}
\renewcommand{\thefigure}{S\arabic{figure}}
\setcounter{figure}{0}

\begin{figure}[!ht]
\begin{center}
\includegraphics[width = \linewidth]{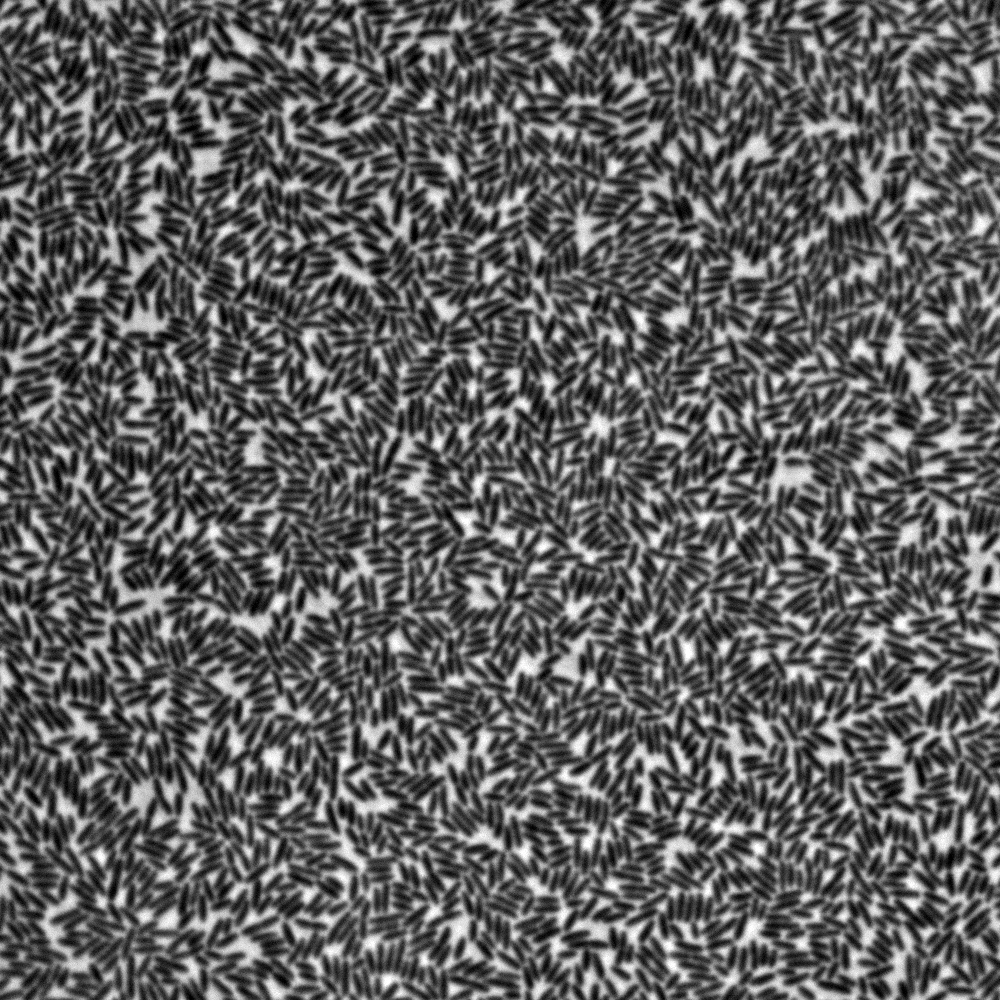}\\
\caption{\textbf{Low density Wild-Type experiment.} $\phi = 0.546$, $\tau_p = 0.57 s$; size: 500x500 pixels; spatial resolution: 0.176 $\mu{}m$/pixel; temporal resolution: 20 frames/second. The full field of view of the movie is presented here but we lowered the spatial and temporal resolution respectively 2-fold and 5-fold to reduce file size.}
\label{simov:WT_low}
\end{center}
\end{figure}

\begin{figure}[!ht]
\begin{center}
\includegraphics[width = \linewidth]{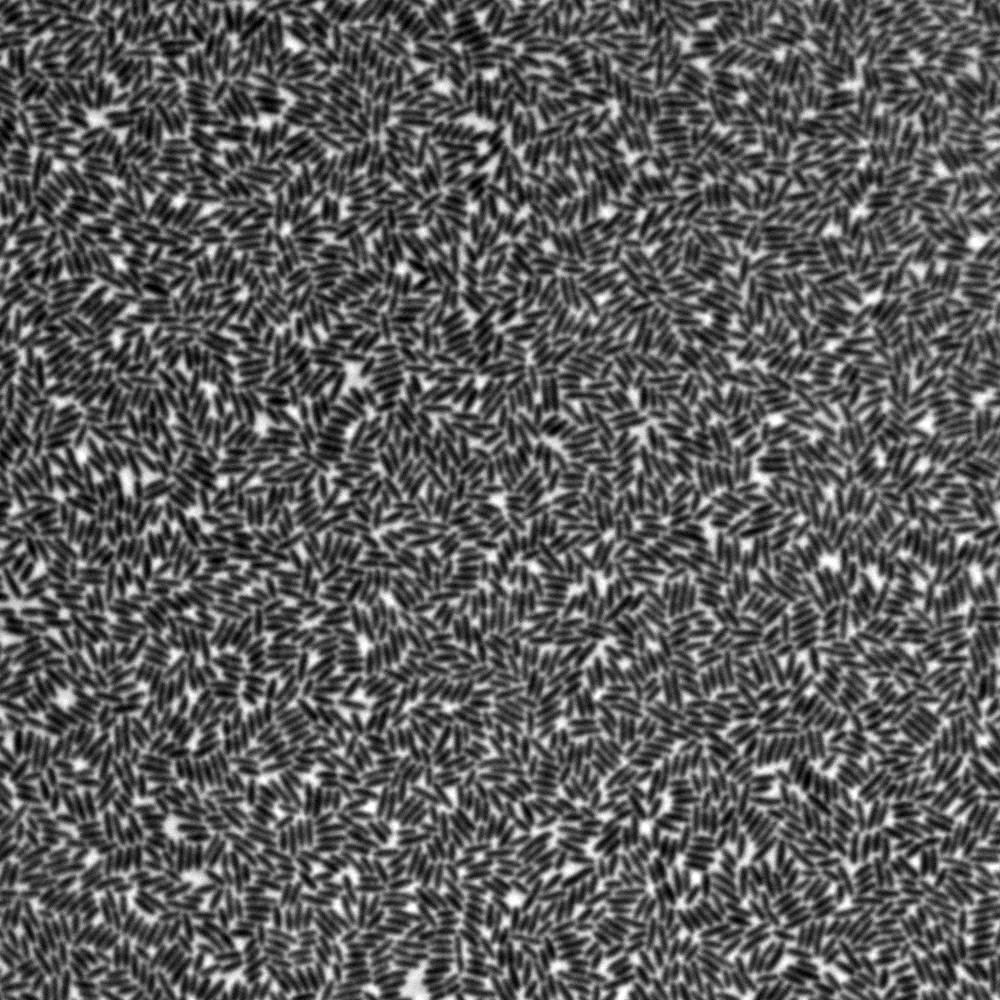}\\
\caption{\textbf{Intermediate density Wild-Type experiment.} $\phi = 0.616$, $\tau_p = 1.45 s$; size: 500x500 pixels; spatial resolution: 0.176 $\mu{}m$/pixel; temporal resolution: 20 frames/second. The full field of view of the movie is presented here but we lowered the spatial and temporal resolution respectively 2-fold and 5-fold to reduce file size.}
\label{simov:WT_mid}
\end{center}
\end{figure}

\begin{figure}[!ht]
\begin{center}
\includegraphics[width = \linewidth]{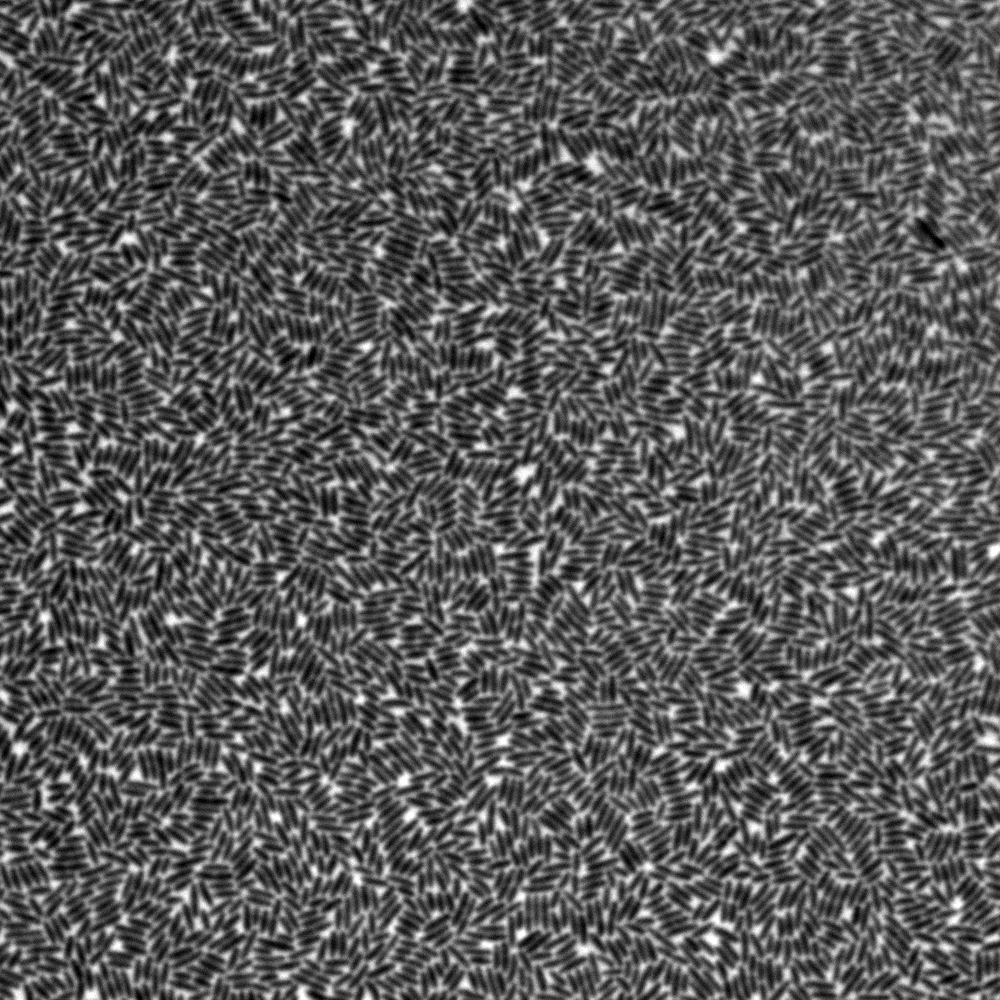}\\
\caption{\textbf{High density Wild-Type experiment.} $\phi = 0.655$, $\tau_p = 3.51 s$; size: 500x500 pixels; spatial resolution: 0.176 $\mu{}m$/pixel; temporal resolution: 20 frames/second. The full field of view of the movie is presented here but we lowered the spatial and temporal resolution respectively 2-fold and 5-fold to reduce file size.}
\label{simov:WT_high}
\end{center}
\end{figure}

\begin{figure}[!ht]
\begin{center}
\includegraphics[width = \linewidth]{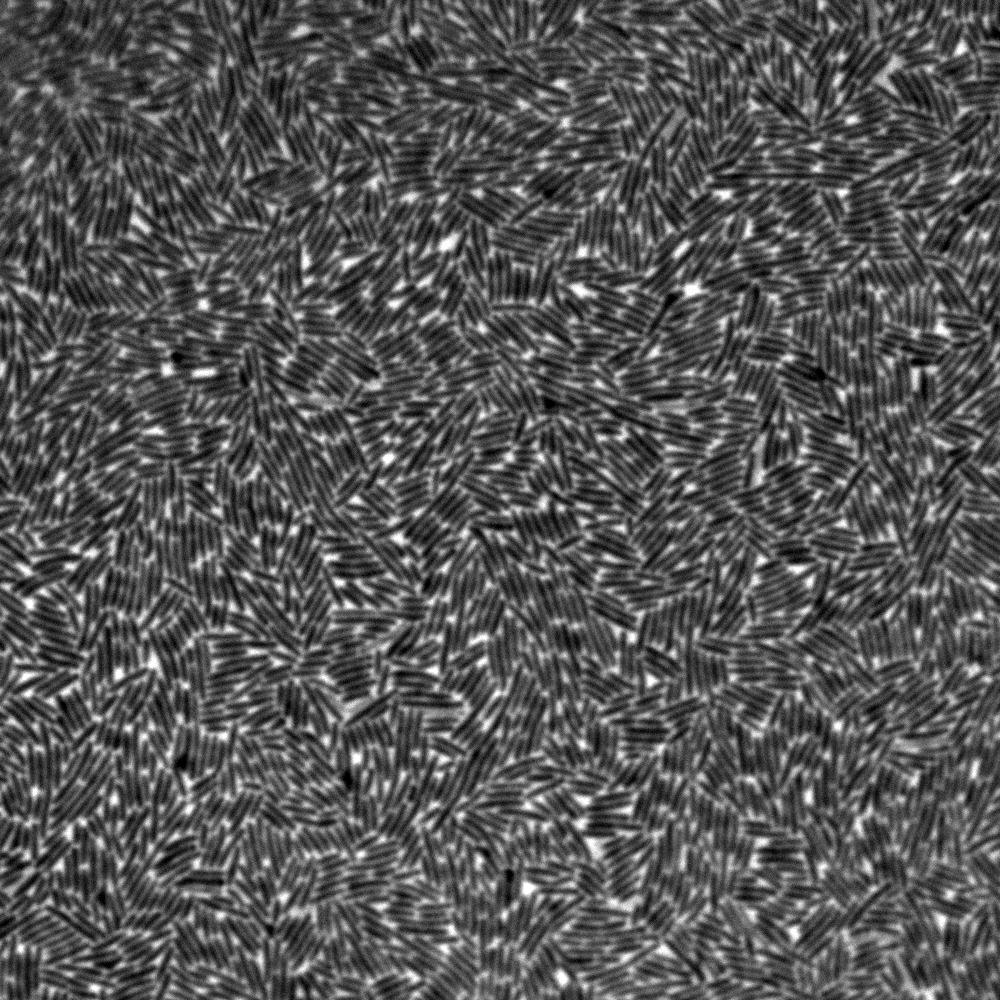}\\
\caption{\textbf{High density Hyperswarmer \textit{frik} experiment.} $\phi = 0.624$, $\tau_p = 356 s$; size: 500x500 pixels; spatial resolution: 0.176 $\mu{}m$/pixel; temporal resolution: 20 frames/second. The full field of view of the movie is presented here but we lowered the spatial and temporal resolution respectively 2-fold and 5-fold to reduce file size.}
\label{simov:Cl5}
\end{center}
\end{figure}

\begin{figure}[!ht]
\begin{center}
\includegraphics[width = \linewidth]{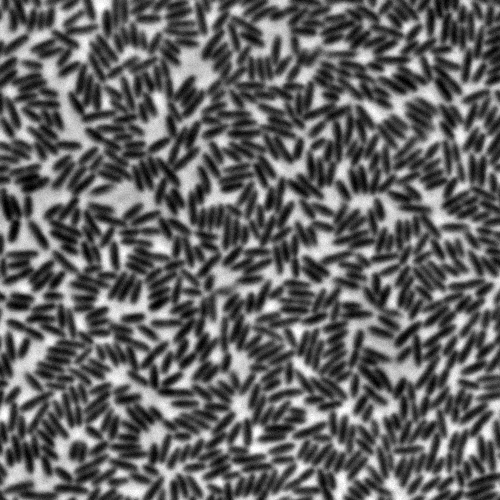}\\
\caption{\textbf{Low density \textit{cheR1} experiment.} $\phi = 0.530$, $\tau_p = 0.07 s$; size: 250x250 pixels; spatial resolution: 0.176 $\mu{}m$/pixel; temporal resolution: 100 frames/second. Because of the high speed of the cells in this movie, the full temporal resolution was needed to observe the cell displacements. To limit the size of the movie file, we then performed a crop of a quarter of the full field of view. We then reduced the spatial resolution 2-fold.}
\label{simov:C1}
\end{center}
\end{figure}

\begin{figure}[!ht]
\begin{center}
\includegraphics[width = \linewidth]{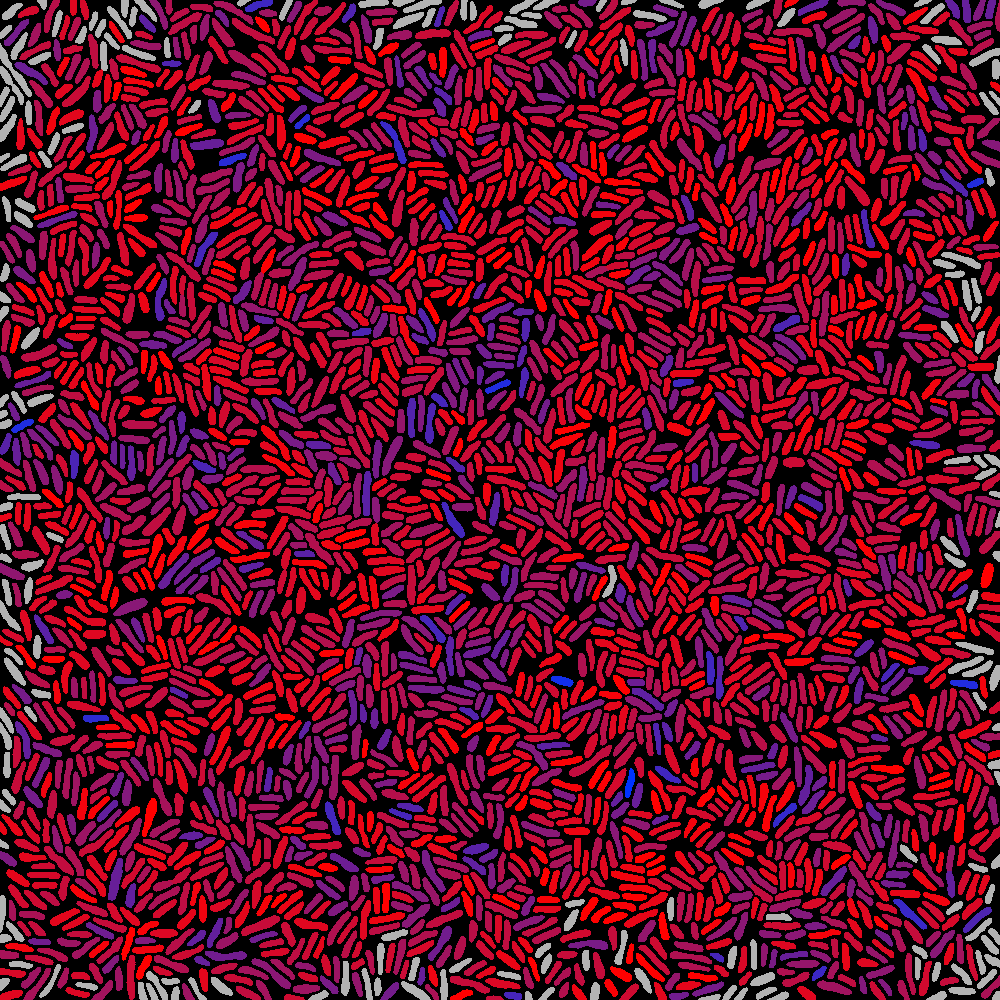}\\
\caption{\textbf{Low density Wild-Type experiment, with cells colored by their squared displacements at $\tau_p$.} $\phi = 0.546$, $\tau_p = 0.57 s$; size: 500x500 pixels; spatial resolution: 0.176 $\mu{}m$/pixel; temporal resolution: 20 frames/second. White cells correspond to cells whose trajectory is shorter than $0.57 s$, making it impossible to compute a squared displacement at $\tau_p$. The full field of view of the movie is presented here but we lowered the spatial and temporal resolution respectively 2-fold and 5-fold to reduce file size.}
\label{simov:WT_low_SD}
\end{center}
\end{figure}

\begin{figure}[!ht]
\begin{center}
\includegraphics[width = \linewidth]{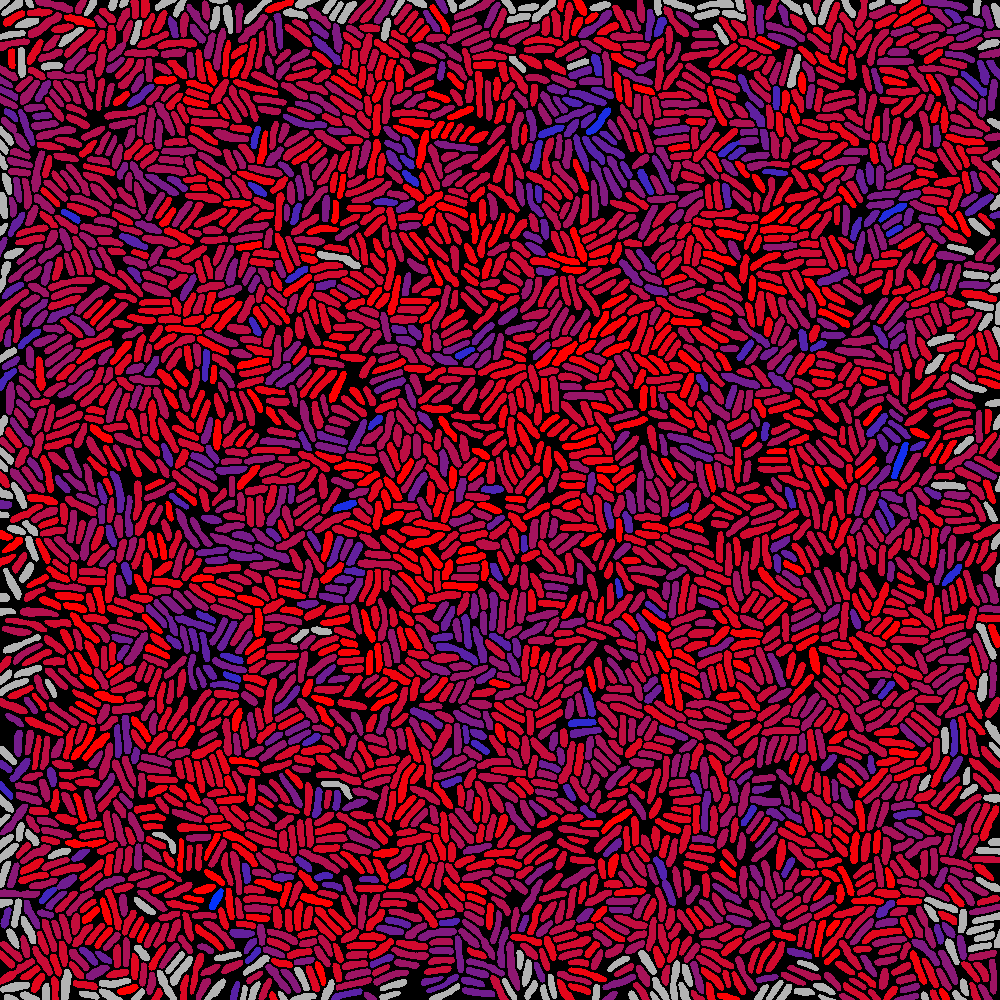}\\
\caption{\textbf{Intermediate density Wild-Type experiment, with cells colored by their squared displacements at $\tau_p$.} $\phi = 0.616$, $\tau_p = 1.45 s$; size: 500x500 pixels; spatial resolution: 0.176 $\mu{}m$/pixel; temporal resolution: 20 frames/second. White cells correspond to cells whose trajectory is shorter than $1.45 s$, making it impossible to compute a squared displacement at $\tau_p$. The full field of view of the movie is presented here but we lowered the spatial and temporal resolution respectively 2-fold and 5-fold to reduce file size.}
\label{simov:WT_mid_SD}
\end{center}
\end{figure}

\begin{figure}[!ht]
\begin{center}
\includegraphics[width = \linewidth]{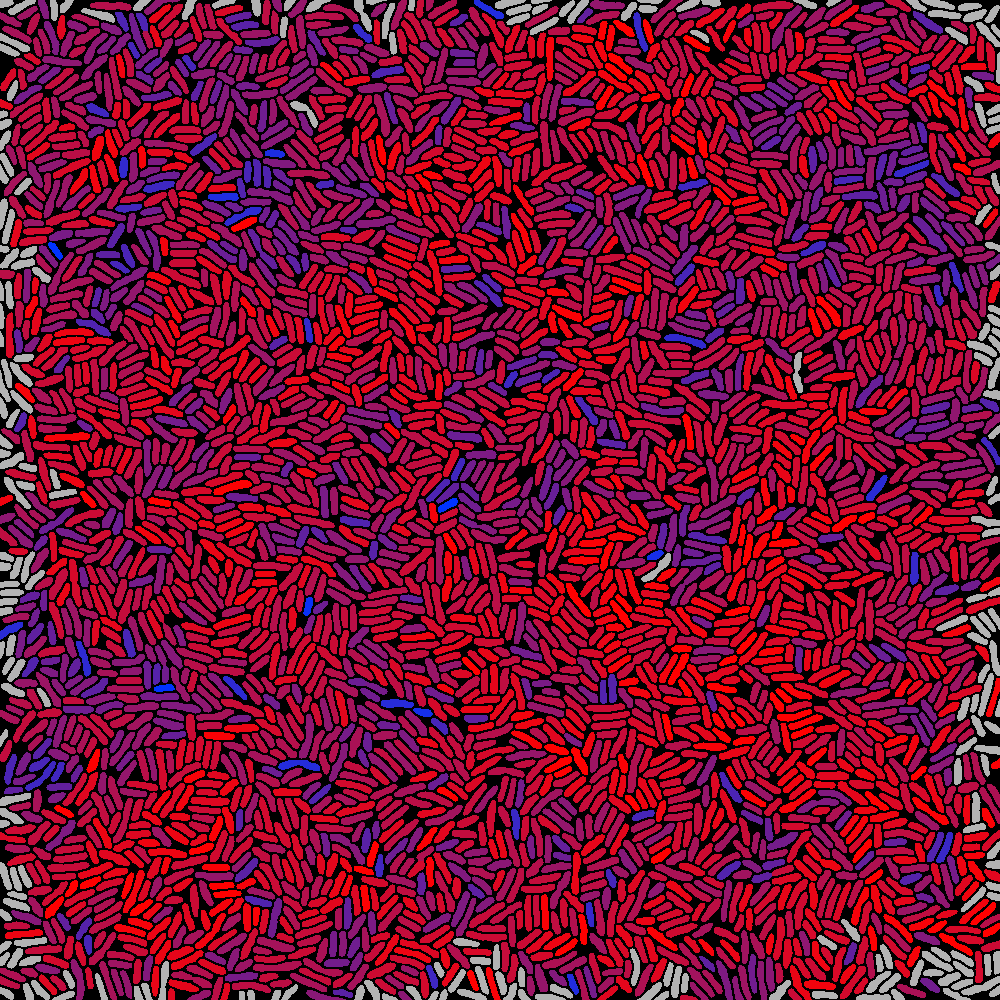}\\
\caption{\textbf{High density Wild-Type experiment, with cells colored by their squared displacements at $\tau_p$.} $\phi = 0.65$, $\tau_p = 3.51 s$; size: 500x500 pixels; spatial resolution: 0.176 $\mu{}m$/pixel; temporal resolution: 20 frames/second. White cells correspond to cells whose trajectory is shorter than $3.51 s$, making it impossible to compute a squared displacement at $\tau_p$. The full field of view of the movie is presented here but we lowered the spatial and temporal resolution respectively 2-fold and 5-fold to reduce file size.}
\label{simov:WT_high_SD}
\end{center}
\end{figure}

\begin{figure}[!ht]
\begin{center}
\includegraphics[width = \linewidth]{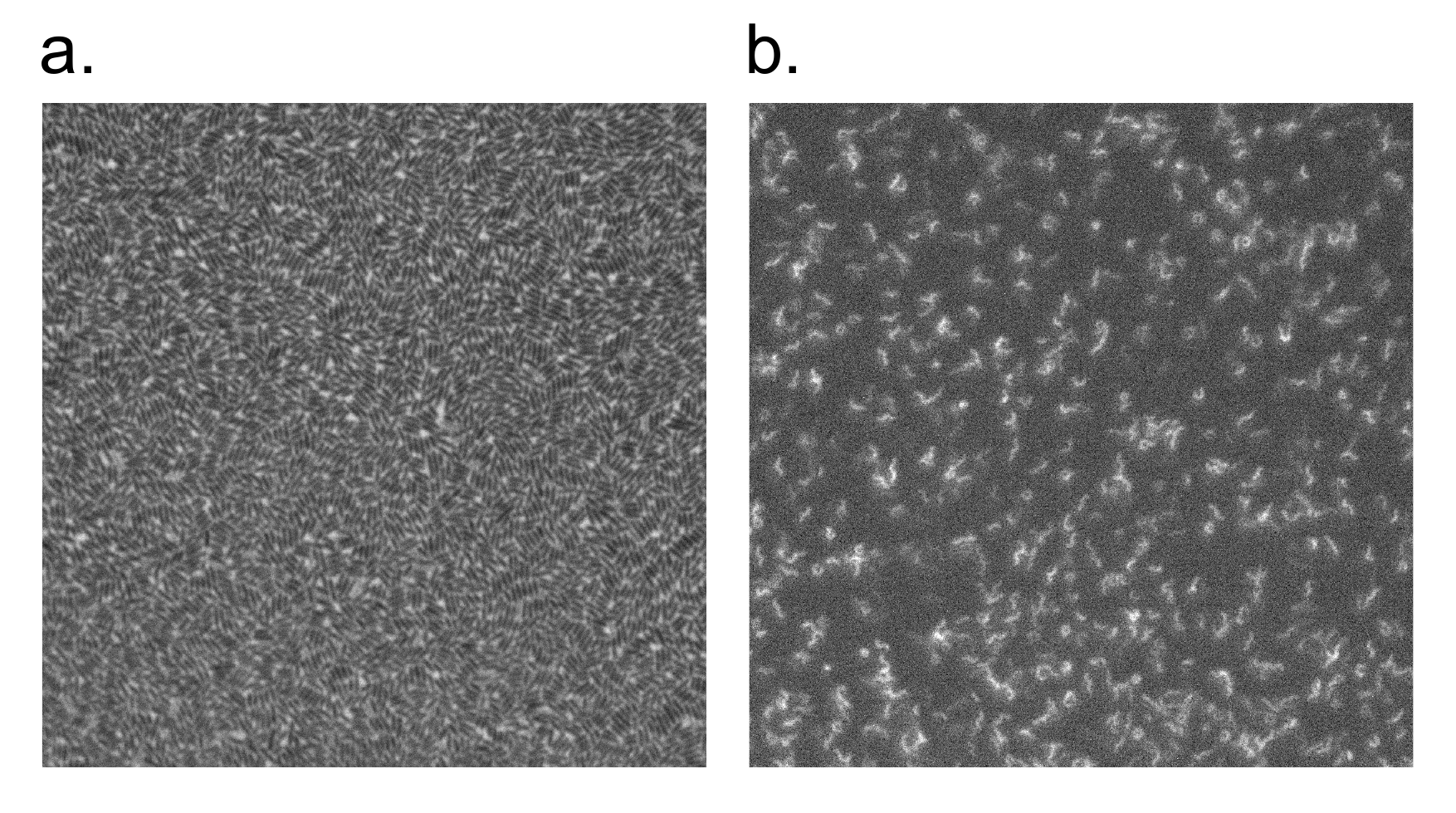}\\
\caption{\textbf{High density \textit{fliC}$^{\text{T394C}}$ experiment with flagella staining.} $\phi = 0.636$; size: 500x500 pixels; spatial resolution: 0.176 $\mu{}m$/pixel; \textbf{(A)} Phase contrast movie to show flagella stained with Alexa-568 maleimide, temporal resolution: 20 frames/second. \textbf{(B)} Fluorescence movie, temporal resolution: 10 frames/second. The movie (B) was acquired immediately after the movie (A).}
\label{simov:FliC}
\end{center}
\end{figure}

\end{document}